\documentclass[12pt,preprint]{aastex}

\author	
{Payam Davoodi,\altaffilmark{1}
Seb Oliver,\altaffilmark{1}
Maria del Carmen Polletta,\altaffilmark{3}
Michael Rowan-Robinson,\altaffilmark{2}
Richard S. Savage,\altaffilmark{1}
Ian Waddington,\altaffilmark{1}
Duncan Farrah,\altaffilmark{4}
Tom Babbedge,\altaffilmark{2}
Carol Lonsdale,\altaffilmark{3,4}
Tracey Evans,\altaffilmark{4}
Fan Fang, \altaffilmark{4}
Eduardo Gonzalez-Solares,\altaffilmark{5}
Tom Jarrett,\altaffilmark{4}
David L. Shupe,\altaffilmark{4} \\
Brian Siana,\altaffilmark{3}
Harding E. Smith,\altaffilmark{3}
Jason Surace\altaffilmark{4} and
C. Kevin Xu\altaffilmark{4}
}

\shorttitle{Modelling the colour distributions of galaxies in SWIRE}
\shortauthors{Davoodi. P., et al.}

\begin{document}

\title{Parametric modelling of the 3.6$\mu$m to 8$\mu$m colour distributions of galaxies in the SWIRE Survey}   

\begin{abstract}

We fit a parametric model comprising a mixture of multi-dimensional
Gaussian functions to the 3.6 to 8$\mu$m colour and optical photometric redshift distribution of  galaxy populations in the ELAIS-N1 and Lockman Fields of the Spitzer Wide-area Infrared Extragalactic Legacy survey (SWIRE).  
For 16,698 sources in ELAIS-N1 we find our data are best modelled (in the sense of the Bayesian Information Criterion) by the sum of four Gaussian distributions or modes ($C_a$, $C_b$, $C_c$ and $C_d$). 

We compare the fit of our empirical model with predictions from
existing semi-analytic and phenomological models. 
We infer that our empirical model provides a better description of the
mid-infrared colour distribution of the SWIRE survey than these
existing models. This colour distribution test is thus a powerful
model discriminator and is entirely complementary to comparisons of
number counts.  

We use our model to provide a galaxy classification scheme and explore
the nature of the galaxies in the different modes of the
model. Population $C_a$ is found to consist of dusty star-forming
systems such as ULIRG's, over a broad redshift range. 
Low redshift late-type spirals are found in population $C_{b}$, where
PAH emission dominates at 8$\mu$m, making these sources very red in
longer wavelength IRAC colours. Population $C_c$ consists of dusty
starburst systems with high levels of star-formation activity at
intermediate redshifts. Low redshift early-type spirals and
ellipticals are found to dominate Population $C_d$. We thus find a
greater variety of galaxy types than one can with optical photometry
alone.

Finally we develop a new technique to identify unusual objects, and
find a selection of outliers with very red IRAC colours.
These objects are not detected in the optical, but have very strong
detections in the mid-infrared. These sources are modelled as dust-enshrouded,
strongly obscured AGN, where the high mid-infrared emission may either be
attributed to dust heated by the AGN or substantial star-formation.
These sources have $z_{ph}\sim2-4$, making them incredibly infrared
luminous, with a $L_{ir} \sim 10^{12.6-14.1}L_{\odot}$.
\end{abstract}


\altaffiltext{1}{\scriptsize{Astronomy Centre, Department of Physics \& Astronomy, University of Sussex, Brighton, BN1 9QH, UK.}}
\altaffiltext{2}{\scriptsize{Astrophysics Group, Blackett Laboratory,
      Imperial College London, Prince Consort Road, London, SW7 2BW, UK.}} 
\altaffiltext{3}{\scriptsize{Centre for Astrophysics \& Space Sciences, University of California, San Diego, La Jolla, CA 92093-0424, USA.}}  
\altaffiltext{4}{\scriptsize{Infrared Processing \& Analysis Centre, California Institute of Technology, 100-22, Pasadena, CA 91125, USA.}}
\altaffiltext{5}{\scriptsize{University of Cambridge, Institute of
    Astronomy, The Observatories, Madingley Road, Cambridge, CB3 0HA, UK.}}

\section{Introduction}

Wide-field survey astronomy is revolutionising astrophysical research
in particular the study of galaxy evolution. Surveys such as SDSS
(York et al.\ 2000), 2dF (Colless 1999), IRAS (Neugebauer et al.\
1984), 2MASS (Kleinmann et al.\ 1994), FIRST (Becker et al.\ 1994),
NVSS (Condon et al.\ 1998) now provide us with a detailed
multi-wavelength picture of millions of galaxies in the local
Universe. The Spitzer Wide-area Infrared Extragalactic Legacy survey
(SWIRE - Lonsdale et al.\ 2003, 2004) now provides us with similarly
detailed sample of galaxies at $z\sim 1$.  These huge and complex data
sets require us to apply a variety of new techniques to extract the
vast wealth of information they contain.

In this paper we explore a parametric technique for studying the
colour distribution of the SWIRE galaxies, providing: a compact
description of the data; a method for source classification; and a
recipe for the identification of outliers.

The statistical properties of galaxies in blank field surveys can be
used to understand galaxy evolution in a number of ways.  The most
basic approach is to explore the surface density of sources as a
function of their flux in a single band and compare this with
models. Such ``number count" analyses have provided clear evidence for
galaxy evolution since early radio surveys (Rowan-Robinson\ 1968).
Spitzer number counts have already revealed strong evolution in
far-infrared bands (e.g. Papovich et al.\ 2004, Dole et al.\ 2004),
and from mid-infrared bands (e.g. Fazio et al.\ 2004).  The SWIRE
number counts will be discussed by Shupe et al.\ (2006) and Surace et
al.\ (2006).  Spectroscopic or photometric redshifts permit a more
direct understanding of the properties of galaxies through, the study
of luminosity functions (e.g. SWIRE luminosity functions - Babbedge et al.\ 2005, Onyett et al.\ 2005).  Analysing the observed-frame
colour distribution of galaxies is a natural extension of the number
count analysis and provides an alternative to photometric redshifts
for exploiting the colour information in extra-galactic surveys. 
Until recently, both models and observations have not been sophisticated
enough to exploit such colour-based techniques. With 
the mid-to-far infrared wavelength coverage of $Spitzer$ (Werner
et al.\ 2004) in seven bands and at sensitivities not previously
encountered in any infrared surveys, we have now reached a level of
maturity where such a colour-based analysis is possible.   

New surveys also bring new challenges in the classification of
sources.  Methods include $\chi^2$ minimisation to fit the spectral
energy distribution (SED) of each source with a library of galaxy
templates, e.g. Bolzonella et al.\ (2000), Farrah et al.\ (2003),
Rowan-Robinson et al.\ (2003). Such template fitting provides both
spectral type and photometric redshift estimates, but success depends
upon the pre-defined templates being representative of the galaxies
under consideration, limiting our confidence as we explore new regions
of parameter space.  More recent techniques based on density
estimation in colour space (Connolly et al.\ 2000) have been
particularly successful in identifying sub-samples of sources, such as
high redshift QSO's in the Sloan Digital Sky Survey (SDSS - York et
al.\ 2000).

In this paper, we apply an efficient and robust technique to model the
colour distribution of sources from the Spitzer Wide-area InfraRed
Extragalactic legacy survey (SWIRE - Lonsdale et al.\ 2003, 2004). We
compare this parametric description of the data with existing
phenomenological and physical galaxy population models.  We use our
model to classify galaxies and interpret these classifications by
comparison with traditional galaxy templates.

We use results from Spitzer (3.6$\mu$m to 24$\mu$m) and optical
U,\textit{g$^\prime$}, \textit{r$^\prime$}, \textit{i$^\prime$}, Z
imaging of the SWIRE fields of ELAIS-N1 and Lockman. We model the
four-dimensional (3 colour, optical photometric redshift) distribution
of galaxies with a mixture of multi-variate Gaussians using an
Expectation Maximization (EM) Algorithm
\citep{bn12000,connolly2000}. Every source is then classified as
belonging to one of these Gaussian ``distributions" or ``modes".  The advantage of this
classification technique over traditional template fitting is that a
direct application of pre-determined galaxy template libraries is not
required, allowing the identification, classification and
characterisation of both existing and new object types.

A colour based analysis of the source characterisation of SWIRE
populations using galaxy template libraries will be discussed in more
detail by Polletta et al.\ (2006b, in prep.), while an analysis of the
spectral energy distributions and photometric redshifts of SWIRE
sources is given by Rowan-Robinson et al.\ (2005).

In $\S$2, we describe the data sets used in our analysis.
$\S$3 gives a detailed description of the classification technique we use to model the colour distribution of SWIRE sources. In $\S$4, we analyse the properties of sources in each mode. We
extend this analysis in $\S$5, using optical/infrared template colours
and star formation rate/stellar mass indicators to determine the type
of sources in each distribution. In $\S$6, we investigate how well our
empirical model can be used to describe simulated data from
theoretical models. In $\S$7, we employ a method based on our
classifications to identify unusual sources in the fields of ELAIS-N1
and Lockman. Discussion and conclusions are presented in $\S$8.

\section{SWIRE}

The Spitzer Wide-area InfraRed Extragalactic survey (SWIRE), the
largest Spitzer Legacy program, is a wide-area imaging survey, mapping
the distribution of spheroids, disks, starbursts and active galactic
nuclei (AGN) and their evolution from $z$$\sim$3 to the current
epoch. The survey covers $\sim$49 square degrees (in 6 high galactic
latitude fields) in all seven Spitzer bands: 3.6, 4.5, 5.8, and
8$\mu$m with IRAC (Fazio et al.\ 2004) and 24, 70 and 160$\mu$m with
MIPS (Rieke et al.\ 2004), detecting $\sim$2.5 million galaxies down
to $f_{3.6{\mu}m}$$\approx$ 3.7$\mu$Jy.

The large area of SWIRE is important to establish statistically
significant population samples over enough cells that we can resolve
the star formation history as a function of epoch and environment,
\textit{i.e.} in the context of structure formation. The large volume
is also essential for finding rare objects and transitory phenomena.

In this paper, we investigate two of the largest SWIRE fields;
ELAIS-N1 and Lockman, covering a total of $\sim$18 square degrees.

\subsection{ELAIS-N1 and Lockman}

The SWIRE ELAIS-N1 field is centred at $16^{h}$$00^{m}$
+$59^{d}$$01^{m}$, with coverage of $\sim$6.5 sq.deg. IRAC
(3.6$\mu$m, 4.5$\mu$m, 5.8$\mu$m and 8$\mu$m)+MIPS (24$\mu$m)
observations were performed in 2004 January and 2004 Feburary.

The average 5$\sigma$ depths of the ELAIS-N1 sample are 5.0, 9.0,
43, 40 and 311$\mu$Jy at 3.6, 4.5, 5.8, 8, and 24$\mu$m respectively
(Surace et al.\ 2004), consistent with the 90$\%$ completeness levels
for source extraction. For ELAIS-N1, optical U, $g^\prime$,
$r^\prime$, $i^\prime$, Z data (complete to $r'$$\sim$23.5) were taken between
1999-2003 using the Isaac Newton Telescope (INT) Wide-Field Survey 
(WFS; McMahon et al.\ 2001, Gonzalez-Solares et al.\ 2005). The 5$\sigma$ limiting
optical depths in Vega are U=23.4, \textit{g$^\prime$}=24.9,
\textit{r$^\prime$}=24.0, \textit{i$^\prime$}=23.2 and Z=21.9.

The SWIRE Lockman field is centred at $10^{h}$$45^{m}$
+$58^{d}$$00^{m}$, with coverage of $\sim$11.5 sq.deg. IRAC
(3.6$\mu$m, 4.5$\mu$m, 5.8$\mu$m and 8$\mu$m)+MIPS (24$\mu$m)
observations for Lockman were performed on 2004 April and 2004 May.
The average 5$\sigma$ depths for the Lockman field are 5.0, 9.0, 43, 40 and
311$\mu$Jy at 3.6, 4.5, 5.8, 8, and 24$\mu$m respectively (Surace et
al.\ 2004), consistent with the 90$\%$ completeness levels for source
extraction. For Lockman, optical U, $g^\prime$, $r^\prime$, $i^\prime$
data (to $r'$$\sim$25) was taken between 2001 May and 2004 March using
the MOSAIC camera on the Mayall 4m telescope at Kitt Peak National
Observatory (Siana et al.\ 2005, in prep.). The 5$\sigma$ limiting
optical depths in Vega are U=25.0, \textit{g$^\prime$}=25.7,
\textit{r$^\prime$}=25.0, \textit{i$^\prime$}=24.0.

For both fields, fluxes were extracted in 5.8$\arcsec$ radius apertures for
IRAC ($\sim$2-3 times the FWHM beam) and 12$\arcsec$ for MIPS, using SExtractor 
(Bertin and Arnouts,\ 1996). At redshift $z$$<$0.2, the median source
FWHM was found to correspond to 2.3 -- 2.4$\arcsec$ in the
optical and 1.5 -- 1.8$\arcsec$ in IRAC. All SWIRE aperture fluxes were then aperture
corrected for wings. The absolute flux calibrations are correct within
roughly 10$\%$ for IRAC and MIPS 24$\mu$m channel data, and were
confirmed by comparison to 2MASS. Further discussion on the
data processing is given by Surace et al.\ (2004, 2005) and Shupe
et al.\ (2006). We use aperture fluxes for our entire sample since the
light of a galaxy is measured consistently through the same aperture in all bands,
thus allowing an unbiased comparison of the galaxy colours in our
sample.  

The optical data sets for ELAIS-N1 and Lockman were processed with the
Cambridge Astronomical Survey Unit's reduction pipeline
\citep{cs2001}. Full analyses of the photometric data in these fields
are reported by Babbedge (2004), Surace et al. (2004) and Siana et
al. (2006, in prep.).

The ELAIS-N1 field contains 411,015 SWIRE sources. 254,693 of these
sources have optical associations in at least one optical band, are
detected with a SNR of at least 5 in one or more IRAC bands, and their 
24$\mu$m associations have a SNR of at least 3. 
The Lockman field contains 681,587 SWIRE sources, of which
255,908 sources have optical associations with the same SNR criteria as
that described for ELAIS-N1. 
A search radius of 1.5'' was used to bandmerge the optical/SWIRE data for both fields. Stars have been
removed from these samples using a star-galaxy separation criterion - see Surace et al.\ (2004). 

We apply an additional constraint to these datasets, and only consider sources
with 5$\sigma$ detections in all four of the IRAC bands (i.e. 3.6$\mu$m,
4.5$\mu$m, 5.8$\mu$m and 8$\mu$m). Applying this criterion reduces our
sample to 29,675 sources for ELAIS-N1 and 34,712 sources for Lockman.

\subsection{SWIRE photometric redshift catalogues}

For both the ELAIS-N1 and Lockman bandmerged datasets, sources with optical
associations have been analysed with a template-fitting photometric
redshift code. This is ImpZ (Babbedge et al.\ 2004), based on the code
of Rowan-Robinson et al.\ (2003) and further updated based on studies
in Rowan-Robinson et al.\ (2005), which uses optical and near-infrared
data to 4.5$\mu$m to fit six optical galaxy templates and two AGN templates.
The criteria for a source being assigned a photometric redshift is at
least 4 detections in optical (in particular $r'$-band)+infrared
(3.6$\mu$m, 4.5$\mu$m) bands, and a reduced $\chi^2$ $<$ 10.  

Therefore, our final sample consists of 13,865 bandmerged
optical/infrared sources in ELAIS-N1 and 8749 bandmerged
optical/infrared sources in Lockman, assigned photometric redshifts
and with detections in all four IRAC bands.  

\section{Modelling the density function in $N$-dimensional space}

A galaxy can be described by a number of parameters such as flux,
colour and redshift, and these data can be represented as points (or
vectors) in an $N$-dimensional parameter space. Parametric modelling
of the distribution function for these points can dramatically reduce
the data volume and simplify comparison with physical or phenomenological models.
 Identifying structures in the distribution function can help us to understand
the different galaxy populations.

\subsection{General Method}

We assume that the $N$-dimensional density function of
galaxies in the sample is composed of a mixture of multi-variate
Gaussian functions. Each one of these Gaussian ``distributions or modes'' may
represent a ``population'' of galaxies with distinct properties.

We use the code from Connolly et al.\ (2000) to fit $n$ data points
in $N$-dimensional space with $m$ Gaussian distributions. This
code uses an Expectation Maximization Algorithm (Connolly et al.\ 2000) for parameter estimation, and a Bayesian Information Criterion
(BIC -- see e.g. Nichol et al.\ 2000) for model selection, i.e. to
decide how many Gaussian distributions are statistically justified by the data.
For each distribution \textit{i}, the code will output mean co-ordinates
\underline{\textsc{{$\mu$}$ _{i}$}} and an \textit{N}x\textit{N}
covariance matrix $\sum_{i}$ (see Appendix). 

We develop this technique to provide a classification scheme for galaxies.
The sum of all $m$ Gaussian modes is a Probability
Density Function, i.e. it describes the probability that a galaxy
selected at random will have a given set of data values
\underline{\textsc{{$x$}}}, and integrates to 1. Each
Gaussian mode thus has a non unit normalisation
which encapsulates the probability that a galaxy drawn at random from
the population as a whole came from that particular population. We
will use the term Probability Density Function ($PDF$ - see Appendix)
to describe each Gaussian mode. 
These $PDF$s allow us to determine the relative probability that any
galaxy is drawn from any particular distribution. 

There are different ways to then classify the galaxies. We could
relate each galaxy to the mode which gives that galaxy the highest
probability density.  Considering each galaxy treated as an isolated
case, this technique provides the optimal (maximum likelihood)
classification. However, since the distribution functions overlap this
will not provide the best classification for the population
statistics. Therefore, we choose an alternative in which we assign
each galaxy randomly to one of the modes with probability proportional
to the $PDF$ value at the galaxy's position. In practice we found the
choice of technique made very little difference.

In the following section, we discuss how this technique was applied to
our data set.

\subsection{Specifics to the SWIRE ELAIS N1 data}

We consider 13,865 sources from the ELAIS-N1 bandmerged
optical/infrared catalogue for which photometric redshifts have been
assigned using ImpZ (Babbedge et al.,\ 2004). These sources have
detections in all four IRAC bands. 
We use these IRAC bands to produce the following three infrared colour
variables; log($f_{8}$/$f_{5.8}$), log($f_{5.8}$/$f_{4.5}$),
log($f_{4.5}$/$f_{3.6}$), and include photometric redshift 
($z_{ph}$) as a fourth variable. 
Since optical data has been used to determine photometric
redshifts for our bandmerged catalogues, the use of photo-z as a
variable in our analysis can be considered as a non-linear mapping of
optical colours, where the photo-z encodes some of the useful
information from the optical which we cannot get from infrared 
colours alone. This also means we gain some information on rest-frame 
properties as well as the observed-frame projection. Future work will
include the use of the optical colours individually (Davoodi et al.\ 2006b). 
In addition, the code we are using automatically standardises our variables to
zero mean and unit variance before fitting Gaussian distributions, which removes any dependency on the choice of units (however, the outputs we quote have been renormalised to natural units).

High quality photometry measurements have been taken for over 2.5
million galaxies in the SWIRE Survey - see Surace et al.\ (2004). 
However, the effects of cosmic rays and/or artifacts in imaging
maps can influence photometry measurements for a very small fraction of
sources. This could therefore lead to some galaxies having extreme colours which
would distort any Gaussian distributions fitted to our data.  
Therefore, we first identify such spurious sources algorithmically, by
running the Expectation Maximization Algorithm to fit a
single Gaussian distribution to the entire data set. 
We identify 1$\%$ of sources with the lowest $PDF$ values (139
sources) from our sample. These source all lie far from the main
galaxy locus in colour space, which we also identify by eye using $n$-D visualisation software \renewcommand{\thefootnote}{\fnsymbol{footnote}}XGobi\footnote[1]{\textit{http://www.research.att.com/areas/stat/xgobi/}}.

Analysing the postage stamps of these sources reveals that all have
errors in their photometry. We therefore eliminate all 139
spurious sources from further analysis, reducing our sample down to 13,726 galaxies. 

\section{A parametric description of the SWIRE multi-colour data}

\subsection{Three IRAC Colours plus Photometric Redshift}

Fitting Gaussian distributions to the 13,726 ELAIS-N1 sources in
3-colour-plus-redshift space, we find that the data is best described
by four, four-dimensional Gaussians.

To test the robustness of our fits to the ELAIS-N1 data, we
also fit to sources in the Lockman field (8749 sources). This
analysis gives us a measure of many of the systematic errors in the parameters of our model.
We find the distribution of the data in the Lockman field is best
described by the same four Gaussian distributions that fit the ELAIS-N1 
data set.  The mean redshift and colours of our four distributions for ELAIS-N1
are given in Table 1. We have also assigned errors to these mean
values, based on the variation between the two
SWIRE data sets. The covariance matrix for ELAIS-N1 and errors based
on the covariance matrix are given in Table 6 (see Appendix).
Table 1 also gives an estimate of the expected number of ELAIS-N1 sources
$N_{\rm exp}$ in each of the four modes. $N_{\rm obs}$ is the actual
number of ELAIS-N1 sources according to our $PDF$ classification technique. We find the numbers
determined using our technique are well within 1$\sigma$ of the expected numbers.

Figures 1a-c show IRAC colour-colour projections of sources assigned to
each of the four distributions (labelled $C_a$, $C_b$, $C_c$, $C_d$). Figure 1d
shows the redshift distribution of $C_a$, $C_b$, $C_c$ and $C_d$. We
also illustrate the optical colours of our four distributions (Figures
2a and 2b), although these variables were not used in the fits.

\begin{figure}[!]
\includegraphics[height=18cm,width=18cm]{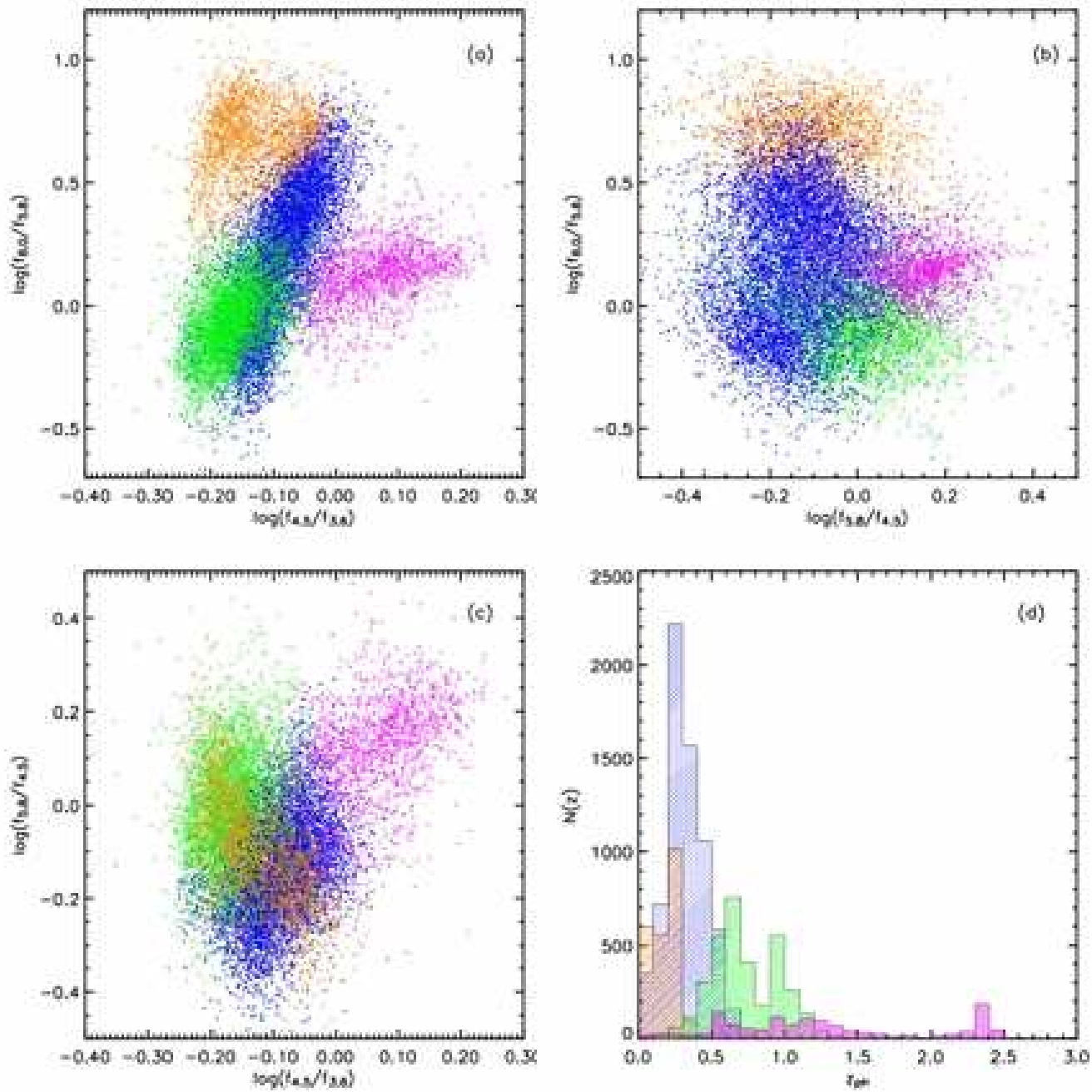}
\caption{\small{Colour and redshift distributions for our sample of
    SWIRE galaxies, classified from mixtures modelling in 3-IRAC
    colour plus redshift space. Four distributions have been identified;
    $C_a$ - magenta, $C_b$ - orange, $C_c$ - green, $C_d$ - blue. (a):
    log($f_{4.5}$/$f_{3.6}$) against log($f_{8.0}$/$f_{5.8}$), (b)
    log($f_{5.8}$/$f_{4.5}$) against log($f_{8.0}$/$f_{5.8}$) and (c)
    log($f_{4.5}$/$f_{3.6}$) against log($f_{5.8}$/$f_{4.5}$) (d)
    Photometric redshift histogram of the four distributions.}}
\end{figure}

\begin{figure}[!]
\includegraphics[height=9cm,width=18cm]{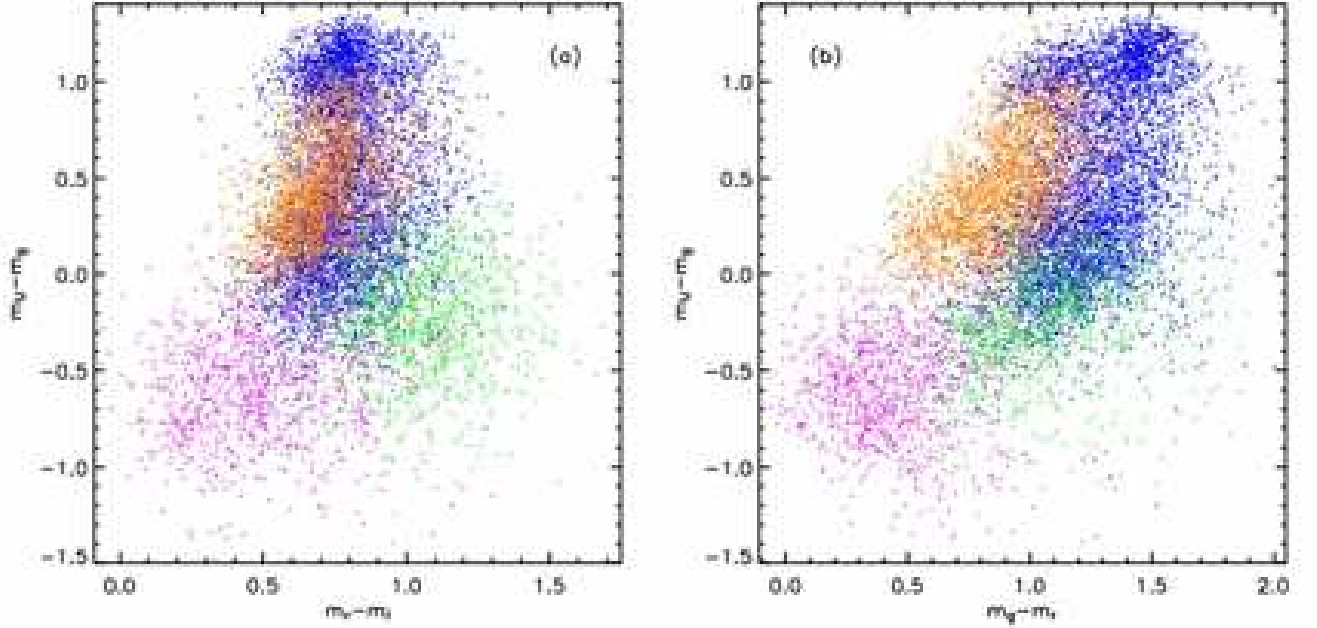}
\caption{\small{Optical colour-colour projections of SWIRE galaxies
    classified in 3-IRAC colour plus redshift space; $C_a$ -
    magenta, $C_b$ - orange, $C_c$ - green, $C_d$ - blue. (a):
    ($m_{r}$-$m_{i}$) against ($m_{U}$-$m_{g}$) and (b) ($m_{g}$-$m_{r}$) against ($m_{U}$-$m_{g}$.)}}
\end{figure}

\begin{deluxetable}{ccccccc}
\tablecaption{\small{Mean co-ordinates of the
four distributions in 3-colour-plus-redshift space for
ELAIS-N1. Errors are estimated from field-to-field variations. The
covariance matrix of each distribution is given in Table 6 - see Appendix.}}
\tablewidth{0pt}
\tablehead{
\colhead{$C_{i}$} & 
\colhead{$N_{\rm exp}\pm1\sigma$} &
\colhead{$N_{\rm obs}$} &
\colhead{log} &
\colhead{log} &
\colhead{log} &
\colhead{$\langle$$z_{ph}$$\rangle$} \\
\colhead{} & 
\colhead{} &
\colhead{} &
\colhead{$\langle$($f_{8}$/$f_{5}$)$\rangle$} &
\colhead{$\langle$($f_{5}$/$f_{4}$)$\rangle$} &
\colhead{$\langle$($f_{4}$/$f_{3}$)$\rangle$} &
\colhead{} \\
}
\startdata
$C_a$ & 1420 $\pm$ 38 & 1402 & 0.13 $\pm$ 0.01 &  0.12 $\pm$ 0.02 & 0.07 $\pm$ 0.02 & 1.28 $\pm$ 0.09 \\ 
$C_b$ & 2506 $\pm$ 50 & 2496 & 0.66 $\pm$ 0.06 & -0.10 $\pm$ 0.02 & -0.10 $\pm$ 0.03 & 0.17 $\pm$ 0.03\\ 
$C_c$ & 3465 $\pm$ 59 & 3515 &-0.05 $\pm$ 0.01 & -0.02 $\pm$ 0.02 & -0.16 $\pm$ 0.01 & 0.73 $\pm$ 0.11\\ 
$C_d$ & 6335 $\pm$ 80 & 6313 & 0.17 $\pm$ 0.05 & -0.14 $\pm$ 0.01 & -0.09 $\pm$ 0.01 & 0.32 $\pm$ 0.04\\  
\enddata
\medskip

\footnotesize{$N_{\rm exp}\pm1\sigma$ is an estimate of
  the expected number of galaxies that represent each of our four distributions
  $C_a$, $C_b$, $C_c$ and $C_d$. $N_{\rm obs}$ is the actual number of galaxies assigned to each
  of the four distributions.} 
\end{deluxetable}

Of the four populations in 3-colour-plus-redshift space, population
$C_a$ (magenta) is found to have the most extreme optical and infrared
colours. This population has a mean redshift of $\langle z_{ph a} \rangle$=1.28,
although its redshift distribution is found to be relatively broad,
spanning the redshift range of our entire sample. Galaxies in this
population have very blue optical colours, with ($m_{U}$-$m_{g}$) $<$ 0.0,
($m_{g}$-$m_{r}$) $<$ 1.0 and ($m_{r}$-$m_{i}$) $<$ 1.0. In
comparison, these galaxies are found to have very red IRAC colours, with
log($f_{8.0}$/$f_{5.8}$), log($f_{5.8}$/$f_{4.5}$) and
log($f_{4.5}$/$f_{3.6}$) $>$ 0.0.

Population $C_b$ (orange) contain sources at low redshift (Figure
1d), with mean redshift $\langle z_{ph b} \rangle$ = 0.17 (Table 1). In the
optical, this population is found to have red ($m_{U}$-$m_{g}$) and
($m_{r}$-$m_{i}$) colours. In the infrared, the colours of $C_b$ are
found to be blue in the shorter IRAC bands (log($f_{4.5}$/$f_{3.6}$)
$<$ 0.0), and then become redder at longer wavelengths
log($f_{8.0}$/$f_{5.8}$) $>$ 0.4.  

Population $C_c$ (green) contains sources at intermediate redshift
($\langle z_{ph c} \rangle$=0.73). This population has ($m_{g}$-$m_{r}$) and
($m_{r}$-$m_{i}$) $>$ 0.8, indicating that these galaxies have red
optical colours. In comparison, this same population is found to be
relatively blue in log($f_{4.5}$/$f_{3.6}$) and
log($f_{8.0}$/$f_{5.8}$) colour.    
  
Populations $C_d$ (blue) also contains sources at low redshift, with
$\langle z_{ph d} \rangle$ = 0.32.  In the optical, this population has 
($m_{U}$-$m_{g}$) $>$ 0.0, similar to population $C_b$, although redder
($m_{g}$-$m_{r}$) colour. In addition, an obvious bi-modality can be
seen in ($m_{U}$-$m_{g}$) colour (Figures 2a and 2b), where this
population is separated into two smaller populations at
($m_{U}$-$m_{g}$)$\sim$0.7. 
At shorter IRAC wavelengths, $C_d$ has blue IRAC colours with
log($f_{4.5}$/$f_{3.6}$) and log($f_{5.8}$/$f_{4.5}$) $<$ 0.0. As we move
to the longer IRAC wavelengths, we find that $C_d$ has a broad range
of log($f_{8.0}$/$f_{5.8}$) colour.


\subsection{Classification of sources not assigned photometric redshifts}

3014 sources from the ELAIS-N1 data set (with detections in all four
IRAC bands) could not be assigned photometric redshifts using the
template fitting photometric redshift code, ImpZ. This
is either due to a source having insufficient detections in the
optical bands used for template fitting, or that no galaxy
template provides a good fit to the SED. 

We now classify sources with no photometric redshifts using these same
four distributions, by marginalising our four-dimensional Gaussian
distributions over redshift. The advantage of this technique
over re-classifying using modes identified in pure 3-colour space is that 
we use the same classes for sources with and without photometric 
redshift and use these classes to give us some idea about the redshift 
distribution of the sources without photometric redshifts.

\begin{figure}[!]
\begin{center}
\includegraphics[height=9cm,width=9cm]{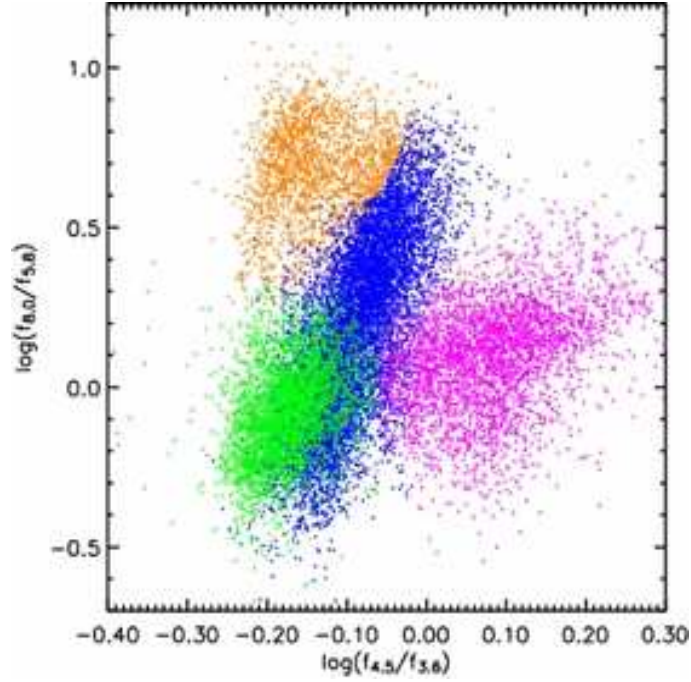}
\end{center}
\caption{\small{log($f_{4.5}$/$f_{3.6}$) against
    log($f_{8.0}$/$f_{5.8}$) colour-colour diagram of 16,740 galaxies. 13,726 of
    these sources have been assigned photometric redshifts, 3014
    sources have not been assigned redshifts. Marginalising our four Gaussian
    distributions in 3-colour-plus-redshift space allows
    sources not assigned photometric redshifts to be classified from 3-colour data; $C_a$ - magenta, $C_b$ - orange, $C_c$ - green, $C_d$ - blue.}}
\end{figure}

\begin{deluxetable}{cccc}
\tablecaption{\small{Number of sources assigned to each of the four marginalised distributions in 3-colour space}}
\tablewidth{0pt}
\tablehead{
\colhead{$C_{j}$} & 
\colhead{$N_{\rm 1} (\%)$} &
\colhead{$N_{\rm 2} (\%)$} &
\colhead{$N_{\rm 3} (\%)$} 
\\
}
\startdata
$C_a$ & 1402 (10$\%$)& 1703 (12$\%$)& 2087 (69$\%$)\\
$C_b$ & 2496 (18$\%$)& 2844 (20$\%$)&  100 ( 4$\%$) \\ 
$C_c$ & 3515 (26$\%$)& 3671 (27$\%$)&  337 (11$\%$)\\
$C_d$ & 6313 (46$\%$)& 5508 (41$\%$)&  490 (16$\%$)\\
\enddata
\medskip

\footnotesize{$N_{i}$ is the number of 
  galaxies in each of the four distributions $C_a$, $C_b$, $C_c$ and $C_d$;\\
  ($i$=1) in 3-colour-plus redshift space,
  ($i$=2) by marginalising over redshift, but for galaxies with redshift information,
  ($i$=3) by marginalising over redshift, but for galaxies without redshift information.}
\end{deluxetable}

\clearpage

We integrate the $PDF$ of each distribution ($P_{4D}(\underline{x})$)
over a redshift interval, defined by the mean redshift ($<$$z_{ph}$$>$)
and standard deviation ($\sigma_{z}$) of each distribution:

$$ {P_{3D}(\underline{x}')} = \int_{-\infty}^{+\infty} {P_{4D}(\underline{x})}\,dx_{4} \approx \int_{<z_{ph}>-5\sigma_{z}}^{<z_{ph}>+5\sigma_{z}} {P_{4D}(\underline{x})}\,dx_{4} $$ 

We first re-classify the 13,726 sources with photometric redshifts (discussed
in $\S$4.1) using the marginalised distributions, and compare
with the original classification. We then classify the 3014 sources
not assigned photometric redshifts using the same marginalised distributions.

Figure 3 is a colour-colour projection of the classification of 16,740 
(13,726+3014) sources. Table 2 shows the number of objects assigned to
each of the four marginalised distributions, and also a comparison of the two classification schemes. We
find reasonable agreement between the two sets of classifications with the numbers in each class changing by less than 5\%. 

We then classify sources without redshifts using the marginalised
  distributions $N_{\rm 3}$ (see Table 2) and find that $\sim$70$\%$
  of sources are assigned to population $C_a$. This population has a
  broad redshift range, and contains sources with very red IRAC
  colours. These sources have not been assigned photometric redshifts
  due to optical detections fainter than our detection limits. This
  would mean the optical+3.6$\mu$m+4.5$\mu$m galaxy templates of
  Rowan-Robinson et al.\ (2005) could not be used to model the SED of
  such sources and therefore determine reliable photometric
  redshifts. These sources will be discussed further in $\S$7.  We
  find that 20$\%$ of sources not assigned photometric redshifts have
  been classified to the two low redshift populations $C_b$ and
  $C_d$. We therefore expect these sources to have redshifts of no
  more than 0.6. The remaining $\sim$11$\%$ of our sample have been
  assigned to population $C_c$ where sources were found to have very
  blue IRAC colours, and $z_{ph}$ = 0.3 - 1.2.

\section{Galaxy templates and SFR/Stellar Mass Indicators}\label{sec:templates}

In order to establish the types of galaxy populating each
distribution, we compare with galaxy templates.  First, we compare the
colours of galaxies classified with our scheme, with the SWIRE galaxy
template library, which has full
optical and infrared coverage. Then we use optical and infrared bands
outside the wavelength range of the IRAC data used in the
classification ($U, r^\prime$ and 24$\mu$m) together with 3.6$\mu$m
data to better understand the properties of our four populations.

The SWIRE optical/infrared template library was
compiled especially for the SWIRE samples and includes the following
spectral types; Ellipticals (three Ellipticals of age 2, 5 and 13 Gyr),
Spirals (Sa, Sb, Sc, Sd), Starbursts (M82 and Arp220), QSO, Seyferts,
ULIRGs and obscured AGN. 

Figure 4 illustrates these galaxy templates and our classifications in
IRAC colour space. Interpreting optical U-band and infrared 24$\mu$m as indicators of
star formation rate (SFR), and optical $r^\prime$-band and infrared 
3.6$\mu$m as indicators of stellar mass, we also illustrate the four
populations in ($m_{U}$-$m_{3.6}$) against ($m_{r^\prime}$-$m_{24}$) 
colour space (Figure 5). We find using this combination of the
four bands gives the best separation of the four populations in
colour-colour space, since each colour is made up of an optical and
infrared band with a large wavelength separation between them. Therefore, each
colour is essentially comparing the optical and infrared SED's of each 
population, which will provide more information in colour-colour space
than using an optical ($m_{U}$-$m_{r^\prime}$) and an 
infrared ($m_{3.6}$-$m_{24}$) colour.

Galaxies in population $C_{a}$ have already been found to occupy a distinct region of IRAC colour space by virtue of their strong, red
continua (Lacy et al.\ 2004). Their blue ($m_{U}$-$m_{3.6}$) colour and red 
($m_{r^\prime}$-$m_{24}$) colour suggests they are dusty systems with high
star formation rates such as ULIRGS (Farrah et al.\ 2001), i.e. $L_{\rm
ir}$$>$$10^{12}$$L_{\odot}$. The template tracks (Figure 4) indicate $C_{a}$
is dominated by AGN and dusty star-forming galaxies. 

Modelling of mid-IR SED's based on ISO spectra (Sajina et al.\ 2005)
suggests galaxies in population $C_{b}$ are dominated by PAH
emission in the IRAC bands, particularly at 8$\mu$m. This would
explain why population $C_{b}$ has very red log($f_{8.0}$/$f_{5.8}$)
colours and is separated from the other three distributions in IRAC
colour space. The ($m_{U}$-$m_{3.6}$) and ($m_{r^\prime}$-$m_{24}$)
colours suggest that galaxies in population $C_{b}$
are star-forming systems. Galaxy templates indicate that this
population is dominated by low redshift late-type spirals.  

The ($m_{U}$-$m_{3.6}$) colour of population $C_{c}$ is found to
be redder than the AGN/dusty star-forming systems in population $C_{a}$. 
The ($m_{r^\prime}$-$m_{24}$) colour of $C_{c}$ is found to be similar
to that of population $C_{a}$, an indication that galaxies in $C_{c}$
are also dusty systems. 
Galaxy templates suggest that this population contains spiral galaxies and dusty
starburst systems at intermediate redshifts, where $L_{ir}$$>$$L_{opt}$.

\begin{figure}[!]
\begin{center}
\includegraphics[height=17cm,width=18cm]{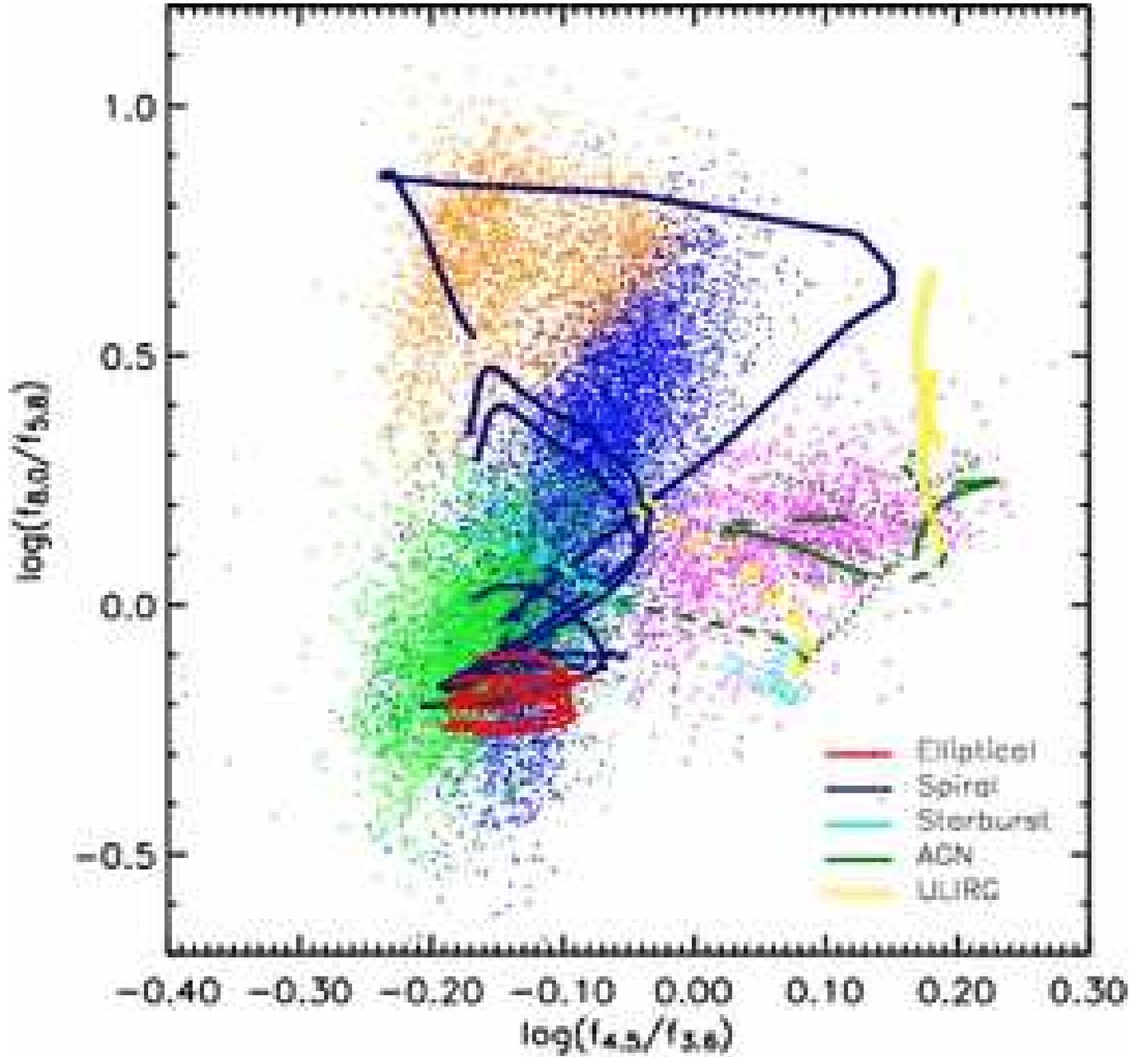}
\caption{\small{log($f_{4.5}$/$f_{3.6}$) against log($f_{8.0}$/$f_{5.8}$) colour-colour diagram of the four galaxy populations ($C_a$ - magenta, $C_b$ - orange, $C_c$ - green, $C_d$) - blue) with the SWIRE galaxy template colours overplotted. Solid curves correspond to $z$$\leq$0.6, dashed curves to $z$=0.6-1.2 and dotted curves $z$$\geq$1.2.}} 
\end{center}
\end{figure}

\begin{figure}[!]
\begin{center}
\includegraphics[height=18cm,width=18cm]{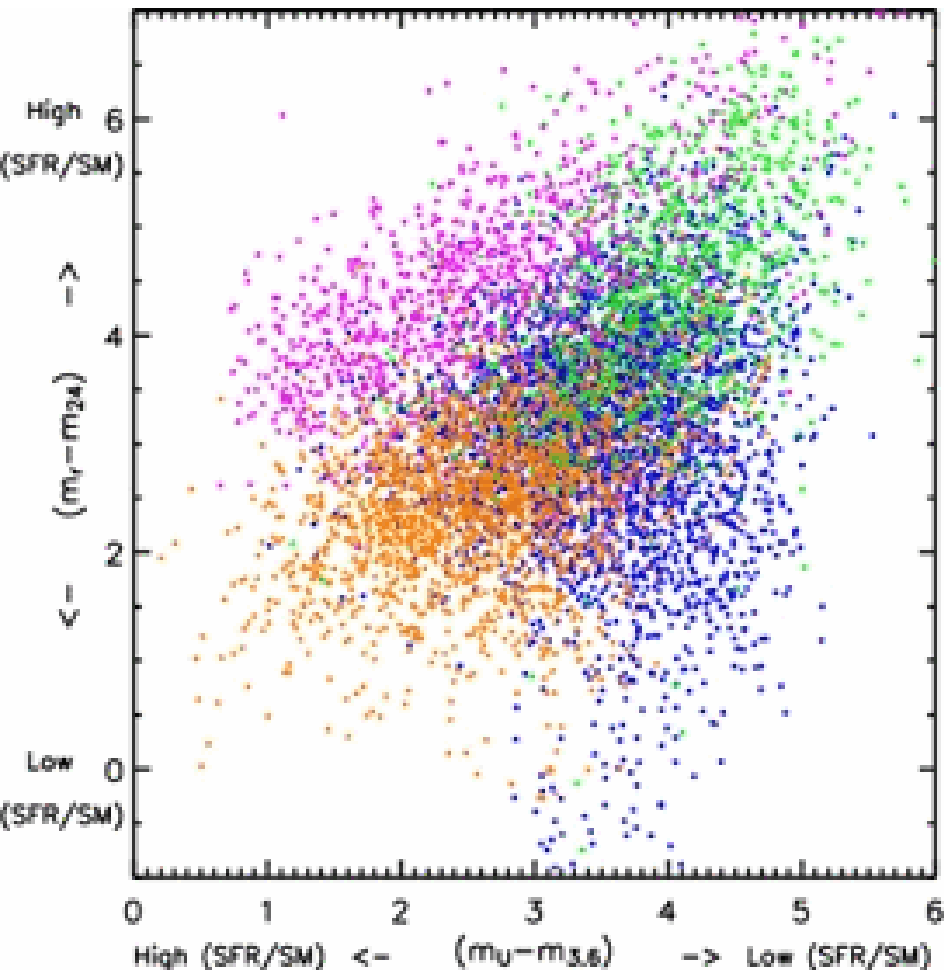}
\caption{\small{($m_{U}$-$m_{3.6}$) against ($m_{r^\prime}$-$m_{24}$) SFR/Stellar mass colour-colour indicators for each of the four populations; $C_a$ - magenta, $C_b$ - orange, $C_c$ - green, $C_d$ - blue.}}
  \end{center}
\end{figure}

$C_{d}$ consists by two sub-populations. Sources with ($m_{r^\prime}$-$m_{24}$)
$>$ 2.5 are found between populations $C_{b}$ and $C_{c}$. These
sources have bluer log($f_{8.0}$/$f_{5.8}$) colour and redder
($m_{r^\prime}$-$m_{24}$) colour than the late-type spirals in
$C_{b}$, and redder log($f_{8.0}$/$f_{5.8}$) colour and bluer
($m_{r^\prime}$-$m_{24}$) colour than dusty systems in $C_{c}$.
They lie in similar regions of IRAC colour space as $C_{b}$,
but their PAH emission is less dominant than $C_{b}$ at 8$\mu$m. Therefore, galaxies in $C_{d}$ with ($m_{r^\prime}$-$m_{24}$) $>$ 2.5 are low redshift early-type spirals. Sources in $C_{d}$ with
($m_{r^\prime}$-$m_{24}$) $<$ 2.5 are low redshift ellipticals. 
These systems have large stellar masses due to their old stellar populations, and their lack of dust content leads to low levels of infrared emission.
The two galaxy classes found within this single mode are separated into two modes in optical studies (e.g. Baldry et al \ 2004). Since our distributions were identified using IRAC colours which at
low redshift detect old stars where ellipticals and early-type spirals
are similar, the low-$z$ elliptical and spiral populations are found
in within this single mode.

\section{Comparing the SWIRE distribution functions to theoretical models}

\begin{figure}[!]
\includegraphics[height=18cm,width=18cm]{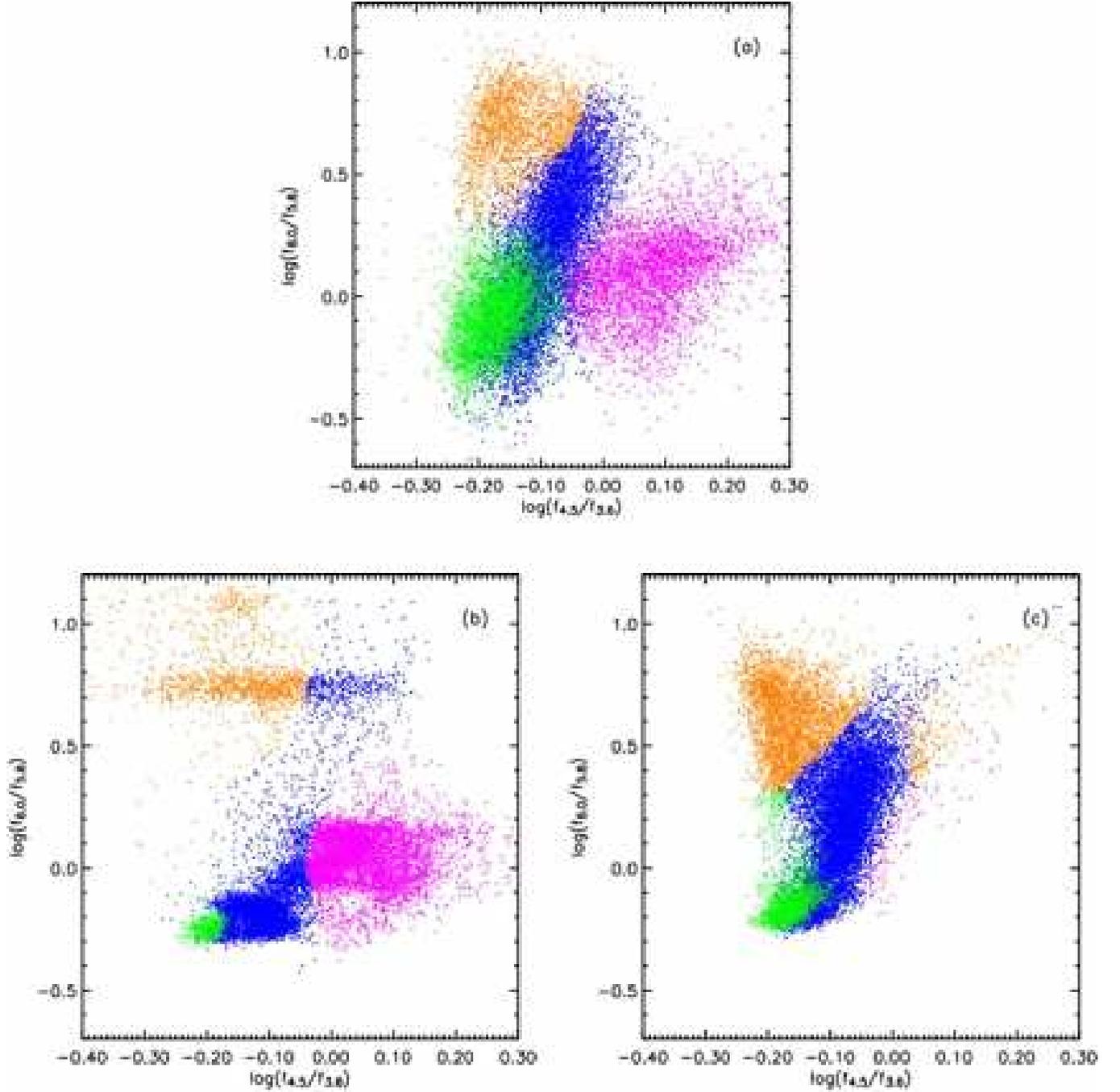}
\caption{\small{log($f_{4.5}$/$f_{3.6}$) against log($f_{8.0}$/$f_{5.8}$) colour-colour distributions of (a) SWIRE, (b)
    Xu and (c) GalICS sources; $C_a$ - magenta, $C_b$ - orange, 
$C_c$ - green, $C_d$ - blue.}}
\end{figure}

\begin{deluxetable}{cccc}
\tablecaption{\small{Comparison of the classification of
    SWIRE, Xu and GalICS sources using our four observed
    distribution functions. $N_{model}$ is the number of sources from
    each model assigned to each distribution function.}}\label{tab:chi2} 

\tablewidth{0pt}
\tablehead{
\colhead{$C_{i}$} & 
\colhead{$N_{\rm SWIRE}$} &
\colhead{$N_{\rm Xu}$} &
\colhead{$N_{\rm GalICS}$} \\
}
\startdata
$C_a$ & 3790 (23$\%$) & 4944 (42$\%$) & 106 (1$\%$)\\ 
$C_b$ & 2944 (17$\%$) & 1218 (11$\%$) & 3167 (22$\%$)\\ 
$C_c$ & 4008 (24$\%$) & 378 (3$\%$) & 1424 (10$\%$)\\
$C_d$ & 5998 (36$\%$) & 5186 (44$\%$) & 9720 (67$\%$)\\  

\enddata
\medskip
\end{deluxetable}

Our parametric description of the SWIRE colour distribution functions
contains a total of 60 parameters - (4 modes, each containing 15 parameters) - see Appendix, Table 6. However, this massively reduced data set still has enormous discriminatory power.

To demonstrate this we fit the distribution functions in
``marginalized" 3-colour-plus-redshift space (our empirical model) to predictions
from one semi-analytic and one phenomological model. These a priori
theoretical models were expected to provide a reasonable description
of the SWIRE observational data.
We use the distribution functions in marginalized 3-colour-plus-redshift space rather than those determined in actual 3-colour-plus-redshift space since the criteria for determining the redshift distributions of the three models presented here will be different, and this may introduce a bias in their comparisons.

First we must determine whether our distribution functions are a good
description of the real data, our previous analysis has merely argued
that these are the best Gaussian mixture description of the data.  We
thus bin our multi-colour data into a multi-dimensional histogram and
apply a $\chi^2$ test. We choose a bin width of 0.05 in log colour space 
and select only those cells that our model predicts will contain two or 
more galaxies. Taking into account the 60 free parameters of the model, 
we determine a reduced $\chi^2\approx 2.7$ -- a surprisingly good fit
given that there is no reason for the underlying distribution of
each population to be exactly Gaussian.

We compare our description of the data with two simulated data sets, a
5 square degree SWIRE mock catalogue from Xu et al.\ (2001, 2003) and a mock
catalogue from GalICS (Hatton et al.\ 2003), made up of five, 1 square
degree observing cones. 

The SWIRE mock catalogue of Xu is based on a set of 
``backward'' galaxy evolution models for nearby infrared bright
galaxies, together with a library of SED's of 837 IRAS 25$\mu$m
selected galaxies. By attaching an appropriate SED from this library
to each source predicted by a given evolution model, a Monte Carlo
algorithm enables simultaneous comparisons of these sources, in a wide
range of wavebands.
The mock catalogue from
GalICS is based on a hybrid model for hierarchical galaxy formation
studies, using the outputs of large cosmological $N$-body simulations
combined with a semi-analytic model. Galaxies in both catalogues have
detections at 3.6$\mu$m, 4.5$\mu$m, 5.8$\mu$m and 8$\mu$m above the
SWIRE 5$\sigma$ flux limits ($\S$2.1).

Since the SWIRE galaxies used to determine our empirical
model have errors associated with their colours, we add random gaussian errors to the colours of sources from 
Xu and GalICS mock catalogues. We do this to 
avoid any bias in the comparison of our empirical model with that of
the theoretical models. The synthetic data consists of 11,726 sources 
from the mock catalogue of Xu and 14,417 sources from GalICS.  
Figures 6a, b and c show IRAC colour-colour projections of SWIRE, Xu
and GalICS sources. It is immediately clear from these plots that the
mock catalogues do not describe the data well.  

To obtain more insight into this we classify the sources in the mock
catalogues by assigning each synthetic source to one of the Gaussian's
as we did with the real data in Section 4.1.  The results are
tabulated in Table 3. We also demonstrate how simple it will be to
compare predictions from future models with our empirical description
of the SWIRE data, repeating the $\chi^2$ analysis to quantify how
well our Gaussian mixtures ``model" fits the synthetic ``data".  

The Xu model was intended to match the monochromatic
number counts, and provides a good description of the IRAC number counts.
However, based on a finite set of ISO templates for the mid-infrared part of
SED's, this model was not designed to match the exact colour
distributions of the IRAC bands. 
Nevertheless it is interesting to explore how it fails in describing
the colour distribution.

Comparing SWIRE (Figure 6a) with Xu (Figure 6b), we find the fit
corresponds to a reduced $\chi^2$$\sim$10$^{19}$, which is mainly
due to the absence of Xu sources in the log($f_{8.0}$/$f_{5.8}$) = 0.2 -- 0.6 colour region of population $C_d$. 
The majority of sources in this population are either concentrated at 
log($f_{8.0}$/$f_{5.8}$)$>$ 0.6 where very low redshift spiral galaxies dominate, or
log($f_{8.0}$/$f_{5.8}$) $<$ 0.2 where we find elliptical galaxies.  The
Xu model is found to severely underpredict redshifted spiral galaxies
(with $z$=0.2 -- 0.5) in the colour region log($f_{8.0}$/$f_{5.8}$) = 0.2 -- 0.6. 
We also find the model of Xu underpredicts spiral and starburst
galaxies with log($f_{8.0}$/$f_{5.8}$) $>$ -0.1 in population $C_c$.

The GalICS model is a more physically motivated model and does
not rely on a fixed set of templates. However, it preceded Spitzer
and has not been tuned to those number counts.  
We find SWIRE sources redder than log($f_{4.5}$/$f_{3.6}$) = 0
have been assigned to class $C_a$ (magenta), making up 23$\%$
of the entire sample. These sources consist of AGN and
dusty systems such as ULIRGS over a broad redshift range (see Section
\ref{sec:templates}).  
However, there are very few GalICS sources in the same region of
colour space (Figure 6c), and population $C_a$ only makes up 1$\%$ of
the GalICS sample.  
The lack of GalICS AGN could therefore account for the high
value of reduced $\chi^2$ ($\sim$10$^{4}$). The GalICS model does not include any
accretion physics which would explain the discrepancy. As a test, we
simulate population $C_a$ to determine whether the absence of these
sources from the GalICS model accounts for the bad fit. We find the
GalICS model still provides a poor description of our observational
data, with a reduced $\chi^2 \approx 51$. However, simulating this
population does show that the absence of population $C_a$ is the main
reason for the inital high value of reduced $\chi^2$. 
As with the model of Xu, the GalICS model also underpredicts spiral and 
starburst galaxies in population $C_c$, with log($f_{8.0}$/$f_{5.8}$) $>$ -0.1. 
However, the GalICS model is found to over-predict galaxies in population $C_b$, with log($f_{4.5}$/$f_{3.6}$) $>$ 0 and log($f_{8.0}$/$f_{5.8}$) $>$ 0.4. Sources assigned to this mode and with these IRAC colours are not found to exist in the observational data set.

This clearly illustrates a catastrophic failure of the a priori models
to match our description of the data. This also demonstrates the need for a complete
model of all extragalactic phenomena when trying to compare models
with observational data (e.g. the addition of an AGN component to GalICS).

\section{Outliers}

Intrinsically rare galaxies and those passing through transient phases
are important for investigating the extreme limits of
galaxy formation and understanding the complete evolutionary behaviour
of galaxies. The search for these sources was a major motivation of
the SWIRE survey and a driver for the wide area and hence large
volume. These sources are likely to have unusual colours and so appear
as ``outliers'' in multi-colour space.

Having modelled the distribution function of galaxies in ELAIS-N1
($\S$4), we use this model to obtain a selection of candidate outliers in
the two SWIRE fields of ELAIS-N1 and Lockman. 
  Our technique for identifying outliers is based on determining
the probability density in each of the four distributions
($PDF_{i}$), which we sum to give a $PDF_{(TOTAL)}$ for every object.  

To identify a sub-sample of outliers, we first need to define an optimal cut
in $PDF_{(TOTAL)}$. We do this by comparing the $PDF_{(TOTAL)}$ of our
observed data with that of a simulated data set modelled 
by our four Gaussians. The simulated data set gives the
expected number of sources $N_{\rm exp}(PDF)$ below 
$PDF_{(TOTAL)}$ (Figure 7a - dashed line), and our SWIRE data set will
give the observed number of sources $N_{obs}$ (Figure 7a - solid
line). Our optimal cut in $PDF_{(TOTAL)}$ will occur in the tail of
the resulting likelihood distributions. 

\begin{figure}[t]
\includegraphics[height=9cm,width=17cm]{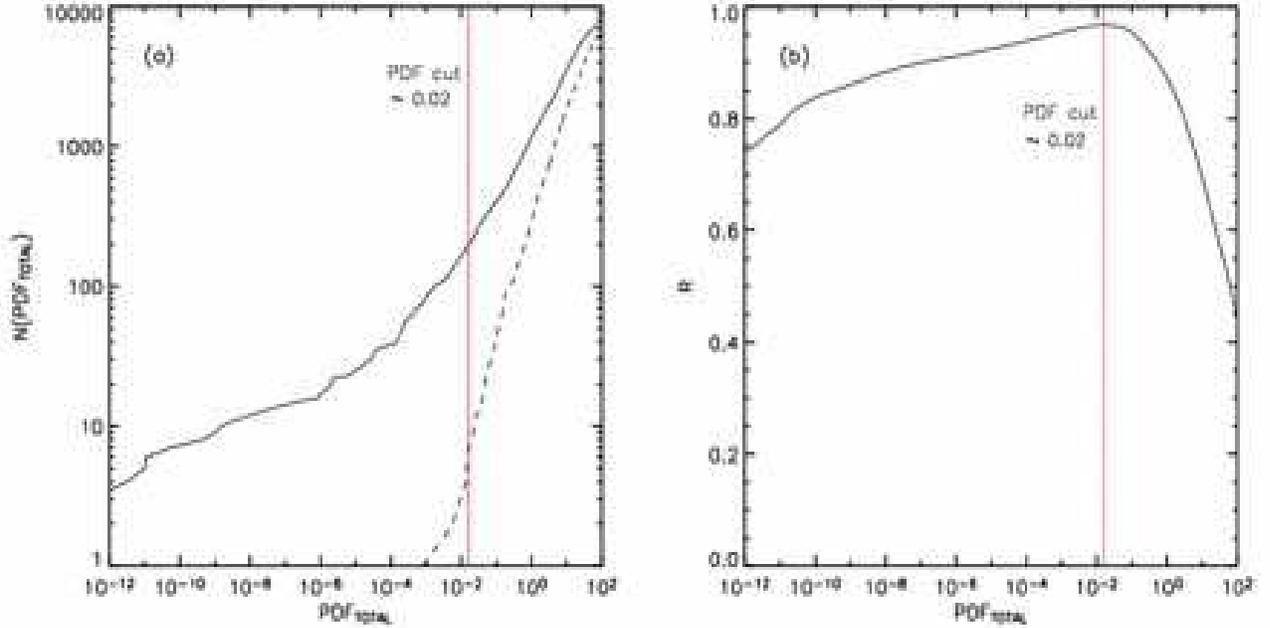}
\caption{\small{(a) Likelihood curve illustrating the number of SWIRE
    sources (solid) as a function of $PDF_{(TOTAL)}$, and the expected
    number of sources (dashed line) based on simulated data modelled
    by Gaussians.
(b) Reliability $R$ as a function of $PDF_{(TOTAL)}$. Sources with
    values of $PDF_{(TOTAL)}$ below the cut (red) are not modelled
    well by our Gaussian distributions. These sources exhibit unusual
    colours compared to the majority of sources in our sample, and so make up our
    sub-sample of candidate outliers.}}  
\end{figure}

\clearpage

We define a reliability, 

$$R = \frac{N_{\rm obs}-N_{\rm exp}}{N_{\rm obs}}, $$

\noindent{being the ratio of unexpected objects to the total number.}

Figure 7b illustrates $R$ as a function of
$PDF_{(TOTAL)}$. $R$ has a maximum of 96\% at
$PDF_{(TOTAL)}$ = 0.022. We therefore apply a cut $PDF_{(TOTAL)}<0.022$ to our sample and identify
242 candidate outliers in ELAIS-N1 and 225 in Lockman.

Since some of these candidate outliers might have spurious detections
in some bands, we use the
IPAC-\renewcommand{\thefootnote}{\fnsymbol{footnote}}Skyview\footnote[2]{\textit{http://www.ipac.caltech.edu/Skyview/}}
software to remeasure their photometry in each of the four IRAC bands,
and compare with their catalogue photometry. This
allows sources which have spurious detections (and therefore
spurious colours) to be eliminated from our candidate list. 
We eliminate 172 sources from ELAIS-N1 and 79 sources from
Lockman. These sources either have bad detections in at least one of
the IRAC bands, or are near very bright sources causing a bias
in their detections. We have thus identified 70 genuine
outliers in ELAIS-N1 and 146 outliers in Lockman. 

\subsection{Candidate outliers}


\begin{figure}[!]
\includegraphics[height=17cm,width=18cm]{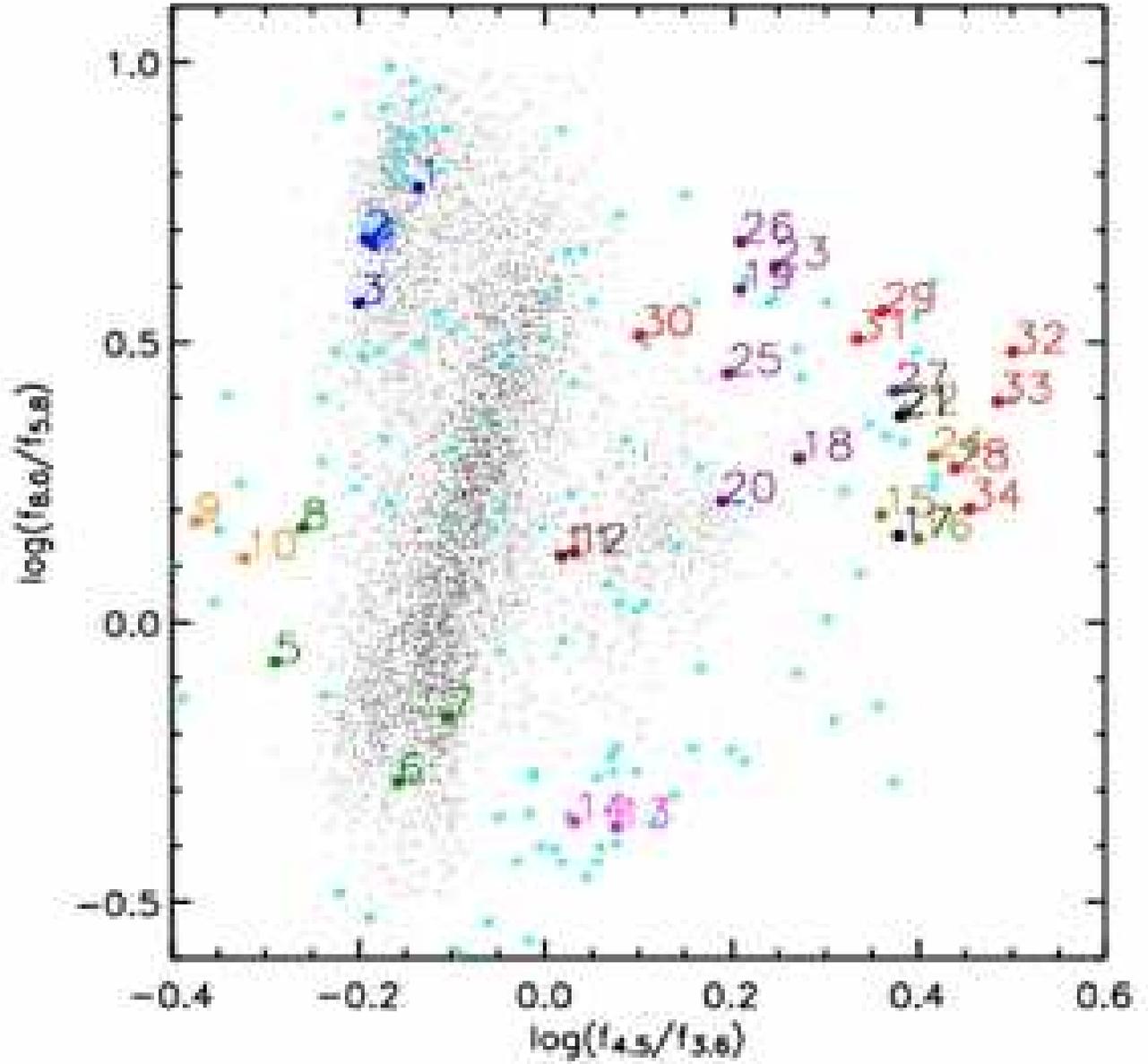}
\caption{\small{log($f_{4.5}$/$f_{3.6}$) against log($f_{8.0}$/$f_{5.8}$) colour-colour projection of 216 outliers
    (cyan). Numbered sources (34 outliers) are representative of the different
    spectral types in each region of colour space; red - Spirals,
    green - Ellipticals, orange - Seyferts, maroon - Starburst/QSO,
    magenta - Spiral/Starburst, brown - QSO, black - Mrk231, purple -
    ULIRG, red - Obscured AGN}}
\end{figure}

Here we discuss a small sub-sample of 34 candidates found in Lockman,
as examples of the types of outliers found in different regions of our
colour-redshift space. Properties of the remaining 182 outliers
from ELAIS-N1 and Lockman can be found \renewcommand{\thefootnote}{\fnsymbol{footnote}}online\footnote[3]{\textit{http://www.astronomy.sussex.ac.uk/$\sim$payam/outliers/index.html}}.

Figure 8 shows their location in IRAC colour-colour
space. Figure 9 (1-34) shows the SEDs of each of these selected
candidates. Tables 4 and 5 show the various properties of each of these
outliers, having modelled the full SED of each outlier with SWIRE
optical/infrared galaxy templates.
These outliers are, by construction, the most unusual objects found in
our sample. Therefore, our fits may not be that good as we may not
have the right templates in our libraries. However, some galaxies may
have reasonable fits because a standard template produces unusual
colours at very specific redshifts. Future SWIRE papers will present
more detailed modelling of unusual objects (Lonsdale et al.\ 2006, in
prep.).

\clearpage

{\textbf{Outliers 1, 2, 3, 4}} - Outliers identified in this region of
colour space have log($f_{8.0}$/$f_{5.8}$)$>$0.5,
log($f_{4.5}$/$f_{3.6}$)$<$-0.1 and are most likely to be star-forming
galaxies. Their SED's are all well represented by a spiral template at
$z_{ph}$=0.05, with low infrared to optical flux ratios. Sources
identified as outliers in this region of colour space have very sharp
PAH features at 8$\mu$m, which gives them peculiar colours at this
specific redshift.

{\textbf{Outliers 5, 6, 7, 8}} - These sources represent outliers with
log($f_{4.5}$/$f_{3.6}$)$<$-0.1 and
log($f_{8.0}$/$f_{5.8}$)$<$0.2. The template colours in $\S$5 suggest
ellipticals dominate this region of colour space. Fitting templates to
the SED's of outliers in this region also suggests that these are
ellipticals. Outliers 5 and 8 which are very blue in
log($f_{4.5}$/$f_{3.6}$) are found to have low infrared to optical
flux ratios and have $z_{ph}$$\sim$0.05. In comparison, outliers 6 and
7 are less blue in log($f_{4.5}$/$f_{3.6}$), yet much bluer in
log($f_{8.0}$/$f_{5.8}$), and have higher redshifts of
$z_{ph}$$\sim$0.2 - 0.5. These sources are not intrinsically unusual
objects but are outliers due to their redshift.

{\textbf{Outliers 9, 10}} - Outliers 9 and 10 are found to have
log($f_{4.5}$/$f_{3.6}$)$<$-0.3, and are identified as outliers due to
PAH features giving odd colours at specific redshifts. With high
infrared to optical flux ratios, such sources are likely to be dusty
systems. The SED's of these outliers are found to resemble that of a
Seyfert 2 galaxy, with intermediate redshifts of $z_{ph}$$\sim$0.8.

{\textbf{Outlier 11}} - This source is found to have
log($f_{5.8}$/$f_{4.5}$)$>$0.5, log($f_{8.0}$/$f_{5.8}$)$>$0.1, and
low infrared to optical flux ratios. With a high detection at
24$\mu$m, the best fitting template to the SED of this object is an
M82 Starburst at $z_{ph}$$\sim$0.07.

{\textbf{Outlier 12}} - This outlier is not found to have extreme infrared colours, but is of particularly high redshift for sources found in this region of colour space. With low infrared to optical flux ratios, the SED of  this outlier resembles that of a Type 1 QSO at a redshift of $z_{ph}$$\sim$2.2.

{\textbf{Outliers 13, 14}} - These source are found to have
log($f_{8.0}$/$f_{5.8}$)$<$-0.3, and represent a cluster of outliers
in this region of colour space. Outlier 13 has high optical and IRAC
emission, but is not detected at 24$\mu$m.  The SED of this outlier
resembles that of a highly luminous spiral galaxy at
$z_{ph}$$\sim$2.4. Outlier 14 is detected at 24$\mu$m, and has high
infrared to optical flux ratios. Therefore, its SED is modelled as
NGC6090 Starburst at $z_{ph}$$\sim$1.4.

{\textbf{Outliers 15, 16, 22, 24}} - These sources are found to have
log($f_{4.5}$/$f_{3.6}$)$>$0.35. The infrared SED's of all four
outliers are very similar, yet outliers 15 and 16 are blank in the
optical. Outliers 22 and 24 have very faint $g'$, $r'$ and $i'$-band
detections, with magnitudes in the range 23 -- 24.4. All these outliers
are expect to have high levels of dust obscuration.  Modelled as Type
1 QSO's with high infrared to optical flux ratios, they have a
redshift range of $z_{ph}$$\sim$1.8 -- 2.4.

{\textbf{Outliers 17, 21}} - Located in similar regions of colour space
as outliers 15, 16, 22 and 24, these two sources are modelled as
Mrk231 (dust-dominated AGN). Both outliers have high 24$\mu$m
detections (hence, high levels of star-formation) characteristic of
Mrk231, and 17 is blank in the optical. At $z_{ph}$$\sim$2, these
sources would have $L_{ir}$$\sim$$10^{13}$-$10^{13.4}$$L_{\odot}$, putting
them in the same catagory as Hyperluminous Infrared Galaxies (HLIRG's
- Farrah et al.\ 2004).
These types of sources are among a rare population of galaxies
recently discovered by Spitzer (Houck et al.\ 2005, Yan et al.\ 2005, Lonsdale et al. \
2006, in prep).

{\textbf{Outliers 18, 19, 20, 23, 25, 26, 27}} - Sources with IRAC
colours log($f_{4.5}$/$f_{3.6}$)$>$0.1 and
log($f_{8.0}$/$f_{5.8}$)$>$0.2. These types of sources are outliers
because they have PAH features which throw out the infrared colours at
particular redshifts. With strong detections in the mid-infrared,
particularly at 24$\mu$m, we expect sources in these regions of colour
space to be very dusty star-forming systems, such as ULIRG's. Outliers
25, 26 and 27 are all blank in the optical, an indication of dust
obscuration since these same systems have relatively low redshifts in
the range $z_{ph}$$\sim$0.17 -- 0.22. Outliers 18, 19, 20 and 23 are the
only ULIRG modelled sources with optical detections. Outliers 18
and 19 have redder IRAC colours than outliers 20 and 23 and are
modelled at $z_{ph}$$\sim$0.18. In comparison, 20 and 23 have much
higher redshifts of $z_{ph}$$\sim$1.2 -- 1.8.

{\textbf{Outliers 28, 29, 30, 31, 32, 33 and 34}} - A selection of
outliers we have identified with very red infrared SED's and no
optical detections. These sources have IRAC colours in the range
log($f_{8.0}$/$f_{5.8}$) = 0.2 -- 0.6 and log($f_{4.5}$/$f_{3.6}$) =
0.1 -- 0.5, making them amongst the reddest objects we have in our
sample.  They are found to have faint near-infrared detections, in
comparison to their very high mid-infrared 24$\mu$m emission, ranging
from 0.7 -- 3mJy. The spectral energy distribution of these sources are
best modelled by dust-enshrouded strongly obscured AGN, where the high mid-infrared emission could either be due to dust heated by the AGN, or substantial star-formation.
At redshifts in the range $z_{ph}$$\sim$ 2 -- 4, their
infrared bolometric luminosities are found to be
$L_{ir}$$\sim$$10^{12.6-14.1}$$L_{\odot}$. If star-formation is
responsible for the high infrared emission in these systems, this
range of infrared bolometric luminosity would correspond to infrared
star-formation rates of 1.5x10$^{3}$ -
5.2x10$^{4}$M$_{\odot}$/yr$^{-1}$. Figure 10 shows the optical and
infrared postage stamps of outlier 28, and illustrates how these
sources are heavily obscured in the optical, but have very high
emission in the mid-infrared bands. 

\clearpage

\subsection{The Number Density of Outliers modelled as dust-enshrouded strongly obscured AGN}

We determine the integrated number density - total of 1/$V_{\rm max}$, of
outliers we have modelled as strongly obscured AGN (outliers 28 -- 34 in $\S$7.1). From our outlier
sample of 216 galaxies from Lockman and
ELAIS-N1, we have identified a total of 12 outliers with SED's corresponding to
that of obscured AGN. Since our sub-sample has $z_{ph}$$\sim$2 -- 4,
we determine the number density of these galaxies within this redshift interval.

$V_{\rm max}$ is the volume corresponding to the maximum redshift at
which a source could be detected by the survey. We set this maximum
redshift by considering a flux-limited sample based on the
mid-infrared 24$\mu$m limit (24$\mu$m$_{lim}$ = 311$\mu$Jy). 
All 12 outliers are found to have 24$\mu$m detections that far exceed
this limit.

We therefore find that these obscured systems have a number density of
$\sim$10$^{-10}$ h$^{3}$ Mpc$^{-3}$ in the high redshift universe,
corresponding to $\sim$0.1$\%$ of the number density of sub-millimetre
galaxy populations (SMG's) identified by SCUBA at z$\sim$2.5 (Chary \&
Elbaz 2001, Chapman et al.\ 2005).

\begin{deluxetable}{c c l l c c c c c c c c c}

\tabletypesize{\scriptsize}
\rotate
\tablewidth{0pc}
\tablecolumns{13}

\vspace{-3.0cm}
\tablecaption{Photometry of 34
  outliers. Columns 5-8 show optical U,
  \textit{g$^\prime$},\textit{r$^\prime$} and \textit{i$^\prime$}
  mag, and 9-13 show IRAC (3.6$\mu$m,
  4.5$\mu$m, 5.8$\mu$m, 8$\mu$m) + MIPS 24$\mu$m in $\mu$Jy.}

\tablehead{
\colhead{Outlier} & 
\colhead{Name} & 
\colhead{RA[J2000]} & 
\colhead{DEC[J2000]} & 
\colhead{U} & 
\colhead{$g'$} & 
\colhead{$r'$} & 
\colhead{$i'$} & 
\colhead{3.6$\mu$m} & 
\colhead{4.5$\mu$m} & 
\colhead{5.8$\mu$m} & 
\colhead{8$\mu$m} & 
\colhead{24$\mu$m}\\

\colhead{ }& 
\colhead{SWIRE$_{-}$J} & 
\colhead{(h:m:s)} & 
\colhead{(d:m:s)} & 
\colhead{(AB)} & 
\colhead{(AB)} & 
\colhead{(AB)} & 
\colhead{(AB)} &
\colhead{($\mu$Jy)} & 
\colhead{($\mu$Jy)} & 
\colhead{($\mu$Jy)} & 
\colhead{($\mu$Jy)} & 
\colhead{($\mu$Jy)}
}

\startdata
 1 & 104752.65+572844.6 & 10:47:52.65 &  +57:28:44.6 & ... & 21.22 & 20.18 & 19.59 & 73.4$\pm$1.0 & 53.6$\pm$1.1 & 44.6$\pm$4.8 & 267.1$\pm$5.6 & 239.2$\pm$34.6 \\
 2 & 105918.35+575042.1 & 10:59:18.35 &  +57:50:42.1 & ... & 20.51 & 19.88 & 19.26 & 56.5$\pm$0.9 & 36.2$\pm$1.3 & 34.1$\pm$4.7 & 165.1$\pm$6.5 & 423.4$\pm$33.4 \\
 3 & 104447.50+575343.8 & 10:44:47.50 &  +57:53:43.8 & ... & 20.92 & 20.16 & 19.67 & 40.1$\pm$0.8 & 25.3$\pm$1.0 & 49.1$\pm$5.1 & 182.5$\pm$5.4 & ... \\
 4 & 105635.36+583244.8 & 10:56:35.36 &  +58:32:44.8 & ... & 20.56 & 19.50 & 18.94 & 98.5$\pm$1.0 & 64.4$\pm$1.2 & 79.1$\pm$4.4 & 373.4$\pm$5.7 & 513.7$\pm$38.4  \\
 5 & 104514.97+580852.4 & 10:45:14.97 &  +58:08:52.4 & ... & 19.59 & 18.28 & 17.59 & 630.8$\pm$4.2 & 323.9$\pm$4.2 & 63.9$\pm$5.3 & 54.3$\pm$5.2 & ... \\
 6 & 104755.39+584814.2 & 10:47:55.39 &  +58:48:14.2 & 23.94 & 22.46 & 20.44 & 19.51 & 112.8$\pm$0.9 & 78.4$\pm$1.0 & 68.8$\pm$3.6 & 35.7$\pm$3.8 & ... \\
 7 & 104628.32+585228.7 & 10:46:28.32 &  +58:52:28.7 & 23.29 & 22.02 & 20.08 & 19.32 & 103.4$\pm$0.9 & 81.2$\pm$1.1 & 44.6$\pm$2.9 & 30.2$\pm$3.6 & ... \\
 8 & 105210.33+582027.5 & 10:52:10.33 &  +58:20:27.5 & ... & 19.89 & 18.35 & 17.65 & 174.0$\pm$1.6 & 95.6$\pm$1.6 & 33.6$\pm$4.9 & 49.5$\pm$5.5 & ...  \\
 9 & 103913.43+594112.1 & 10:39:13.43 &  +59:41:12.1 & ... & ... & 23.13 & 22.05 & 231.3$\pm$2.3 & 97.6$\pm$2.3 & 128.7$\pm$7.1 & 194.6$\pm$9.4 & 891.6$\pm$20.6 \\
10 & 104057.84+581808.2 & 10:40:57.84 &  +58:18:08.2 & ... & ... & ... & 20.66 & 175.5$\pm$1.5 & 83.4$\pm$1.3 & 74.7$\pm$4.5 & 97.2$\pm$5.3 & 633.5$\pm$17.5 \\
11 & 104055.13+580206.6 & 10:40:55.13 &  +58:02:06.6 & ... & 21.60 & 20.66 & 20.45 & 12.0$\pm$0.7 & 12.5$\pm$0.9 & 43.1$\pm$4.9 & 56.7$\pm$5.3 & 1489.8$\pm$15.7 \\
12 & 104807.13+584224.4 & 10:48:07.13 &  +58:42:24.4 & 20.58 & 21.10 & 20.51 & 19.97 & 62.7$\pm$0.8 & 67.3$\pm$0.9 & 80.3$\pm$3.9 & 106.9$\pm$3.9 & 742.4$\pm$15.7 \\
13 & 104330.68+584928.9 & 10:43:30.68 &  +58:49:28.9 & 23.99 & 24.04 & 23.32 & 22.74 & 35.4$\pm$0.8 & 42.2$\pm$1.0 & 46.1$\pm$4.2 & 19.8$\pm$3.9 & ...  \\
14 & 105008.02+574801.5 & 10:50:08.02 &  +57:48:01.5 & ... & 24.51 & ... & 22.76 & 77.1$\pm$0.9 & 82.7$\pm$1.0 & 77.0$\pm$5.1 & 33.9$\pm$5.1 & 295.0$\pm$19.5 \\
15 & 105928.00+572640.4 & 10:59:28.00 &  +57:26:40.4 & ... & ... & ... & ... & 9.8$\pm$0.5 & 22.5$\pm$0.8 & 40.3$\pm$3.8 & 62.7$\pm$4.5 & 238.1$\pm$15.7 \\
16 & 105834.93+574725.3 & 10:58:34.93 &  +57:47:25.3 & ... & ... & ... & ... & 27.0$\pm$0.9 & 67.9$\pm$1.3 & 128.9$\pm$6.3 & 181.6$\pm$6.0 & 516.2$\pm$17.3 \\
17 & 104659.42+584624.0 & 10:46:59.42 &  +58:46:24.0 & ... & ... & ... & ... & 21.4$\pm$0.7 & 51.2$\pm$1.1 & 87.8$\pm$4.2 & 125.4$\pm$4.7 & 607.7$\pm$14.2 \\
18 & 103944.02+573639.1 & 10:39:44.02 &  +57:36:39.1 & ... & 22.67 & 22.05 & 21.53 & 9.5$\pm$0.7 & 17.7$\pm$1.2 & 77.9$\pm$6.1 & 152.7$\pm$6.5 & 1152.0$\pm$19.9 \\
19 & 105700.41+583313.2 & 10:57:00.41 &  +58:33:13.2 & ... & 24.42 & 23.17 & 22.80 & 4.2$\pm$0.5 & 6.9$\pm$0.9 & 15.7$\pm$3.6 & 62.1$\pm$5.1 & 409.8$\pm$19.5 \\
20 & 105844.93+574145.9 & 10:58:44.93 &  +57:41:45.9 & ... & 23.50 & 23.00 & 22.53 & 38.8$\pm$1.0 & 60.1$\pm$1.4 & 30.5$\pm$5.7 & 50.3$\pm$6.3 & 648.8$\pm$18.2 \\
21 & 103752.16+575048.7 & 10:37:52.16 &  +57:50:48.7 & ... & ... & 24.04 & 23.14 & 110.9$\pm$1.5 & 266.4$\pm$1.9 & 566.2$\pm$8.8 & 1321.9$\pm$7.2 & 3727.3$\pm$20.5 \\
22 & 103909.10+580946.0 & 10:39:09.10 &  +58:09:46.0 & ... & 24.00 & 23.64 & 23.06 & 12.1$\pm$0.8 & 29.6$\pm$1.3 & 50.2$\pm$6.2 & 119.2$\pm$6.6 & 396.2$\pm$19.3 \\
23 & 104111.62+582123.0 & 10:41:11.62 &  +58:21:23.0 & ... & ... & ... & 23.06 & 70.3$\pm$1.3 & 124.1$\pm$1.6 & 110.1$\pm$6.4 & 475.2$\pm$6.6 & 1292.9$\pm$17.1 \\
24 & 104000.20+582458.2 & 10:40:00.20 &  +58:24:58.2 & ... & 24.37 & 23.93 & 23.46 & 16.7$\pm$0.9 & 43.4$\pm$1.4 & 82.5$\pm$6.3 & 163.1$\pm$6.7 & 999.6$\pm$17.7 \\
25 & 105817.58+581916.7 & 10:58:17.58 &  +58:19:16.7 & ... & ... & ... & ... & 16.8$\pm$0.7 &  26.5$\pm$1.0 & 84.6$\pm$5.1 & 235.4$\pm$5.6 & 974.6$\pm$19.7 \\
26 & 103508.84+573737.7 & 10:35:08.84 &  +57:37:37.7 & ... & ... & ... & ... & 11.5$\pm$0.8 & 18.6$\pm$1.0 & 43.7$\pm$5.4 & 208.9$\pm$5.7 & 746.6$\pm$20.6 \\
27 & 104148.12+590322.4 & 10:41:48.12 &  +59:03:22.4 & ... & ... & ... & ... & 9.3$\pm$0.7 & 22.0$\pm$0.9 & 41.8$\pm$5.2 & 108.4$\pm$4.8 & 393.6$\pm$20.3 \\
28 & 104314.94+585606.3 & 10:43:14.94 &  +58:56:06.3 & ... & ... &... & ... & 8.1$\pm$0.6 & 22.5$\pm$0.9 & 63.2$\pm$3.6 & 119.0$\pm$4.4 & 965.2$\pm$16.3 \\
29 & 104024.03+571944.4 & 10:40:24.03 &  +57:19:44.4 & ... & ... & ... & ... & 10.8$\pm$0.7 & 25.0$\pm$1.2 & 45.9$\pm$5.8 & 165.0$\pm$6.5 & 723.7$\pm$18.0 \\
30 & 105132.41+591355.3 & 10:51:32.41 &  +59:13:55.3 & ... & ... & ... & ... & 22.0$\pm$0.9 & 27.8$\pm$1.0 & 85.1$\pm$6.3 & 277.0$\pm$5.6 & 2406.3$\pm$18.9 \\
31 & 104132.08+581508.5 & 10:41:32.08 &  +58:15:08.5 & ... & ... & ... & ... & 18.8$\pm$0.9 & 40.6$\pm$1.3 & 85.6$\pm$6.3 & 274.5$\pm$6.9 & 942.8$\pm$19.7 \\
32 & 104931.60+554954.4 & 10:49:31.59 &  +55:49:54.4 & ... & ... & ... & ... & 31.9$\pm$1.1 & 101.0$\pm$1.6 & 309.5$\pm$7.3 & 940.4$\pm$7.9 & 3206.4$\pm$18.0 \\
33 & 104337.28+575830.3 & 10:43:37.28 &  +57:58:30.3 & ... & ... & ... & ... & 3.1$\pm$0.6 & 9.6$\pm$0.9 & 58.9$\pm$4.8 & 146.1$\pm$5.4 & 559.4$\pm$18.3 \\
34 & 103839.02+574533.9 & 10:38:39.01 &  +57:45:33.8 & ... & ... & ... & ... & 28.1$\pm$0.9 & 79.9$\pm$1.1 & 330.4$\pm$7.7 & 526.4$\pm$5.3 & 1617.5$\pm$16.2 \\

\enddata
\end{deluxetable}


\begin{deluxetable}{cccccc}
\tabletypesize{\scriptsize}
\tablecaption{\small{Optical/infrared properties of 34 outliers using SED template fitting.}}
\tablewidth{0pt}
\tablehead{
\colhead{Outlier} &
\colhead{$z_{ph}$} &
\colhead{$PDF$} &
\colhead{Log($L_{opt}$)} & 
\colhead{Log($L_{IR}$)} &  
\colhead{Best Template Fit} \\

\colhead{} & 
\colhead{} &
\colhead{} &
\colhead{$L_{\odot}$} &
\colhead{$L_{\odot}$} &
\colhead{} \\

}
\startdata
 1 & 0.05 & 1.60x10$^{-4}$ & 9.96 & 10.81 & Spiral - Sc \\  
 2 & 0.05 & 3.54x10$^{-5}$ & 9.73 & 10.58 & Spiral - Sc \\  
 3 & 0.05 & 4.11x10$^{-4}$ & 9.59 & 10.45 & Spiral - Sc \\  
 4 & 0.05 & 2.83x10$^{-5}$ & 10.02 & 10.87 & Spiral - Sc \\  
 5 & 0.05 & 2.08x10$^{-6}$ & 11.37 & 7.67 & Elliptical \\    
 6 & 0.49 & 2.40x10$^{-6}$ & 11.86 & 8.52 & Elliptical \\  
 7 & 0.20 & 1.73x10$^{-6}$ & 11.75 & 7.24 & Elliptical \\   
 8 & 0.05 & 8.05x10$^{-3}$ & 10.87 & 7.43 & Elliptical \\   
 9 & 0.84 & 4.02x10$^{-4}$ & 12.47 & 13.32 & Seyfert 2 \\    
10 & 0.83 & 0.01 & 12.42 & 13.27 & Seyfert 2 \\  
11 & 0.07 & 4.66x10$^{-4}$ & 9.02 & 9.87 & M82 Starburst \\  
12 & 2.25 & 2.10x10$^{-5}$ & 12.69 & 12.84 & QSO IR (blue qso) \\ 
13 & 2.41 & 2.94x10$^{-5}$ & 12.31 & 12.38 & Spiral - Sd \\ 
14 & 1.43 & 0.01 & 11.97 & 11.62 & NGC6090 Starburst \\ 
15 & 2.13 & 0.02 & 11.95 & 12.10 & Type 1 QSO \\ 
16 & 1.82 & 1.38x10$^{-3}$ & 12.13 & 12.79 & Type 1 QSO \\ 
17 & 2.14 & 4.43x10$^{-3}$ & 12.20 & 13.34 & Mrk231 \\ 
18 & 0.17 & 3.07x10$^{-5}$ & 9.07 & 10.67 & IRAS22491 \\ 
19 & 0.19 & 8.71x10$^{-7}$ & 8.76 & 10.40 & IRAS22491 \\  
20 & 1.82 & 7.59x10$^{-4}$ & 12.04 & 13.72 & IRAS22491 \\  
21 & 1.82 & 0.01 & 12.70 & 13.05 & Mrk231 \\ 
22 & 2.06 & 7.76x10$^{-3}$ & 12.03 & 12.69 & Type 1 QSO \\ 
23 & 1.26 & 1.53x10$^{-3}$ & 11.82 & 13.19 & IRAS19254 \\  
24 & 2.37 & 1.81x10$^{-3}$ & 12.42 & 13.11 & Type 1 QSO \\ 
25 & 0.22 & 0.02 & 9.27 & 11.13 & IRAS22491 \\  
26 & 0.21 & 1.37x10$^{-3}$ & 9.12 & 10.92 & IRAS22491 \\  
27 & 0.21 & 0.01 & 9.10 & 10.45 & IRAS20551 \\ 
28 & 3.01 & 1.86x10$^{-3}$ & 12.25 & 13.10 & Obscured AGN  \\ 
29 & 2.79 & 5.26x10$^{-3}$ & 12.10 & 12.95 & Obscured AGN \\ 
30 & 4.24 & 2.69x10$^{-4}$ & 13.22 & 14.07 & Obscured AGN \\ 
31 & 1.95 & 0.02 & 11.69 & 12.55 & Obscured AGN \\  
32 & 2.48 & 4.33x10$^{-6}$ & 12.62 & 13.47 & Obscured AGN \\  
33 & 2.90 & 1.34x10$^{-8}$ & 12.09 & 12.94 & Obscured AGN \\  
34 & 2.41 & 5.16x10$^{-6}$ & 12.30 & 13.15 & Obscured AGN \\  

\enddata
\medskip

\footnotesize{Column 2 shows the photometric redshift estimates,
  column 3 gives the sum of the probability density values for each
  outlier, columns 4-5 show optical and infrared luminosity estimates
  based on template fits, and column 6 shows the best-fit templates
  for the SED's of each outlier.}
\end{deluxetable}

\clearpage


\begin{figure}
\begin{center}
\large{\textbf{\textsc{Figure 9 - (1-34)}}}
\end{center}
\includegraphics[height=22cm,width=17cm]{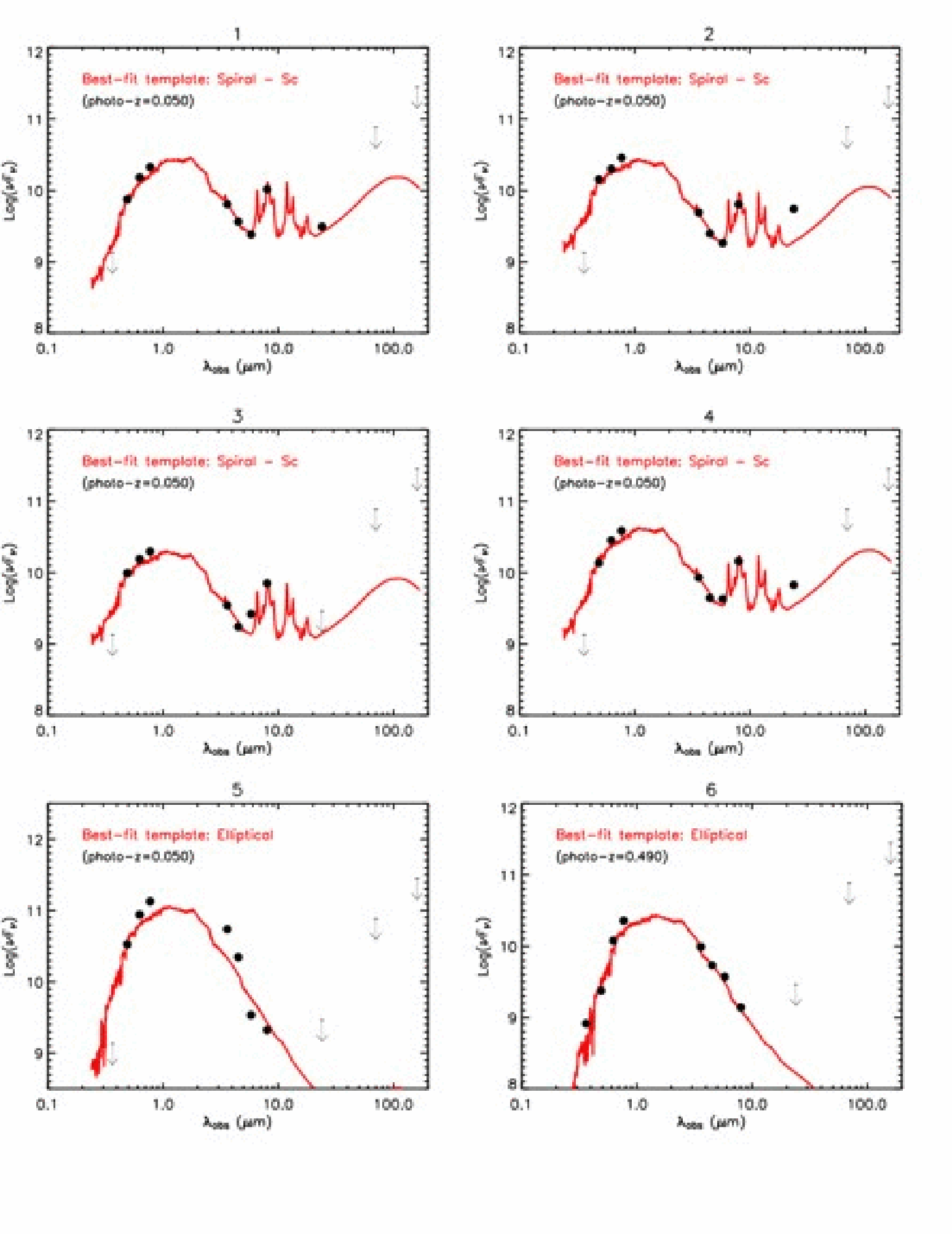}
\caption{\small{Spectral energy distribution (SED) of selected
    outliers modelled using SWIRE galaxy templates.}}
\end{figure}

\clearpage

\begin{figure}

\includegraphics[height=24cm,width=18cm]{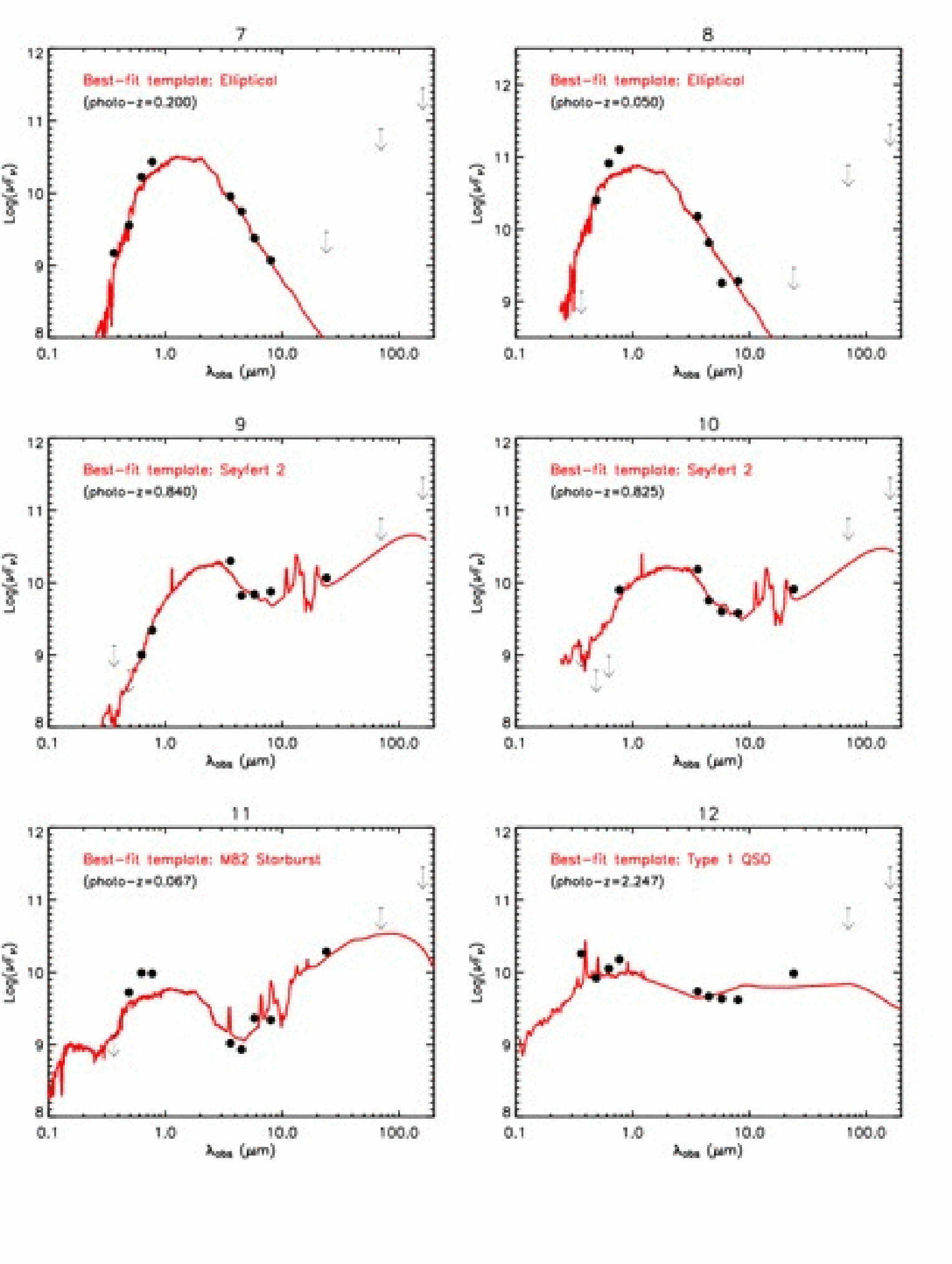}

\end{figure}

\clearpage

\begin{figure}
\includegraphics[height=24cm,width=18cm]{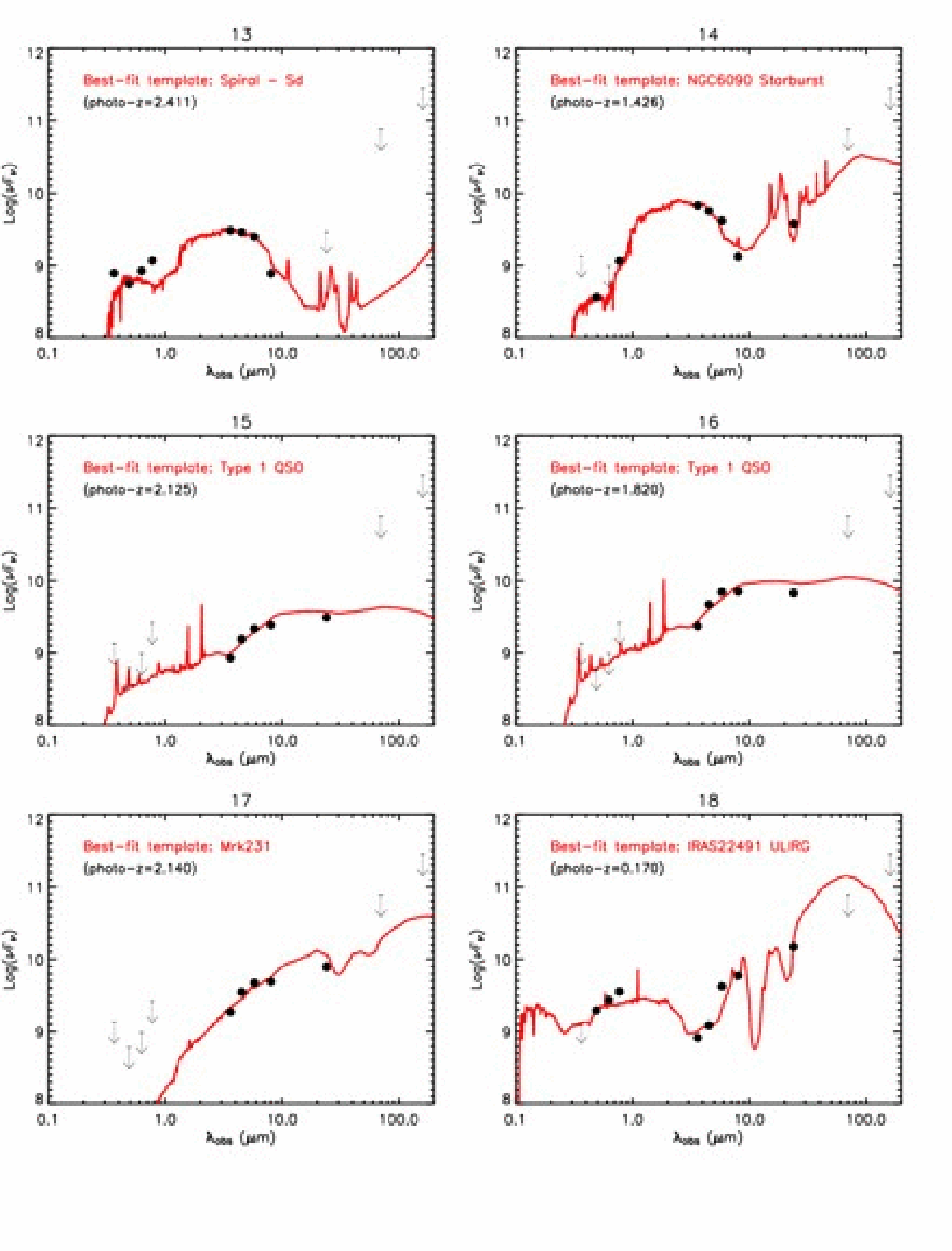}

\end{figure}

\clearpage

\begin{figure}
\includegraphics[height=24cm,width=18cm]{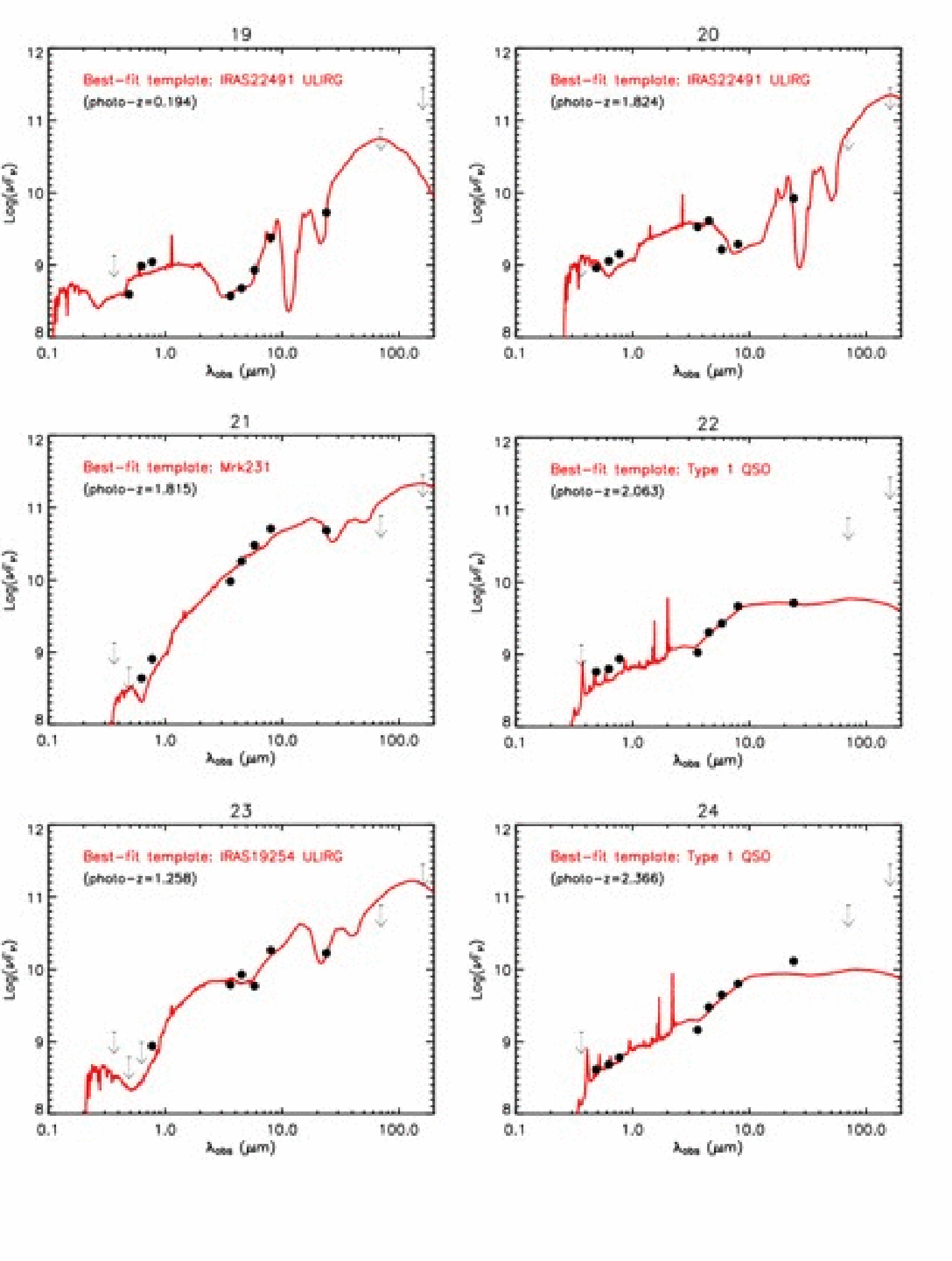}

\end{figure}

\clearpage

\begin{figure}
\begin{center}
\includegraphics[height=18cm,width=18cm]{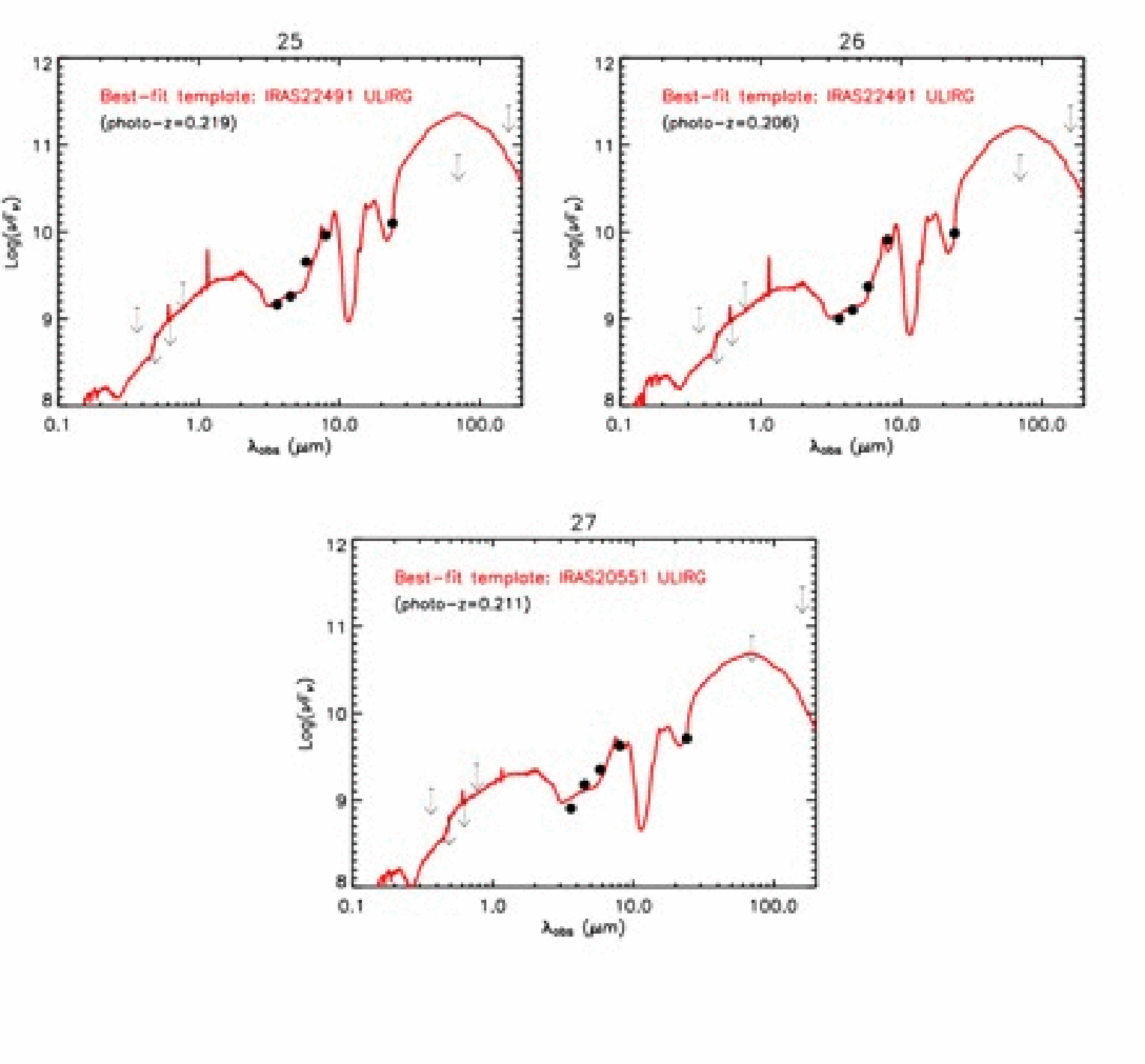}
\end{center}
\end{figure}

\clearpage


\begin{figure}[t]
\begin{center}
\includegraphics[height=20cm,width=15cm]{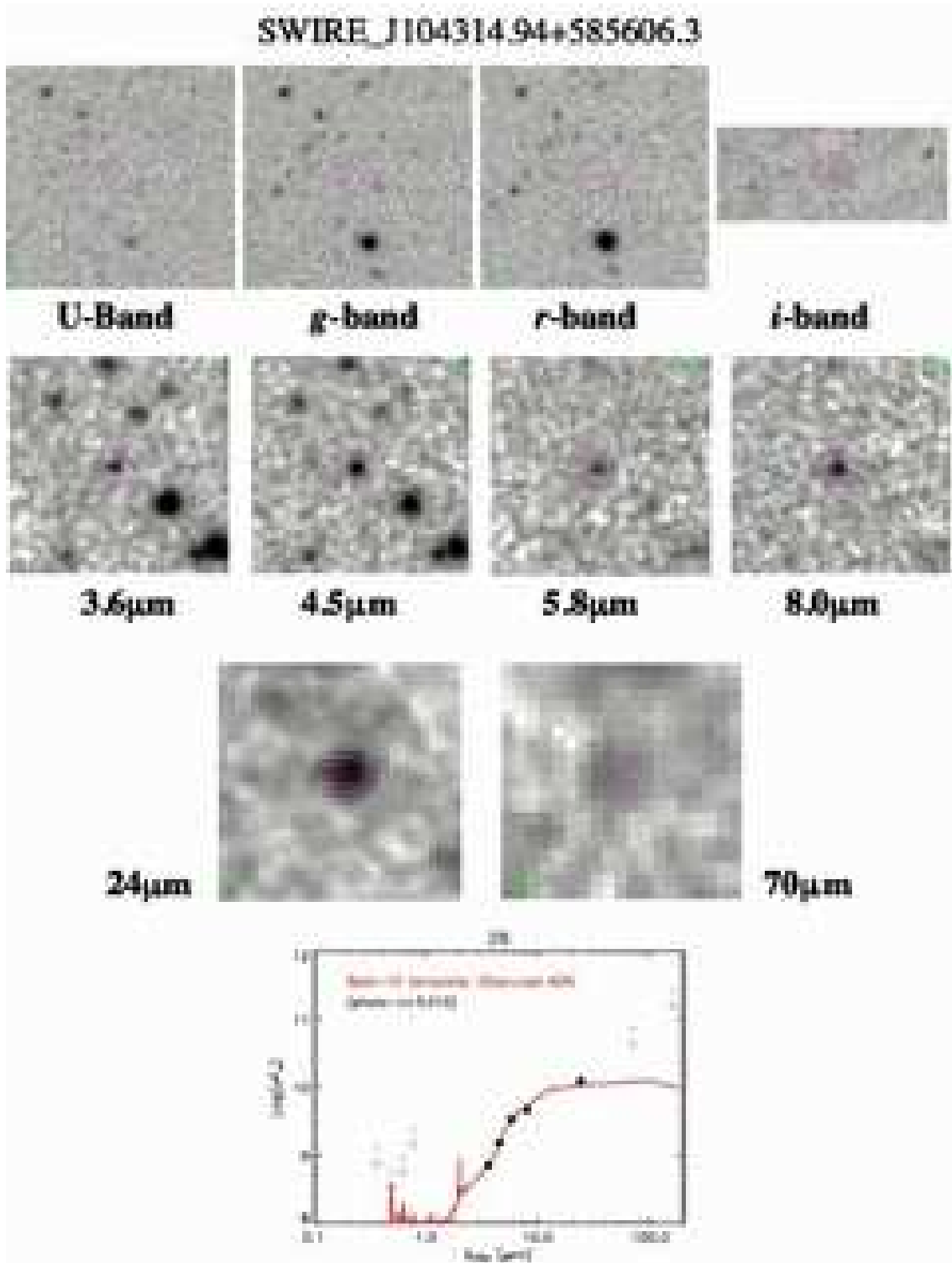}
\caption{\tiny{Fig. 10.- Optical (U, $g'$, $r'$, $i'$) and infrared
  (3.6$\mu$m-70$\mu$m) images of Outlier 28. Also shown is the best
  fitting template to the SED of this object. NOTE: The
  orientation of the optical U, $g'$, $r'$, $i'$ images are offset by
  $\sim$45 degrees clockwise from the infrared 3.6$\mu$m-70$\mu$m images.}}
\end{center}
\end{figure}

\newpage

\begin{figure}
\includegraphics[height=20cm,width=17cm]{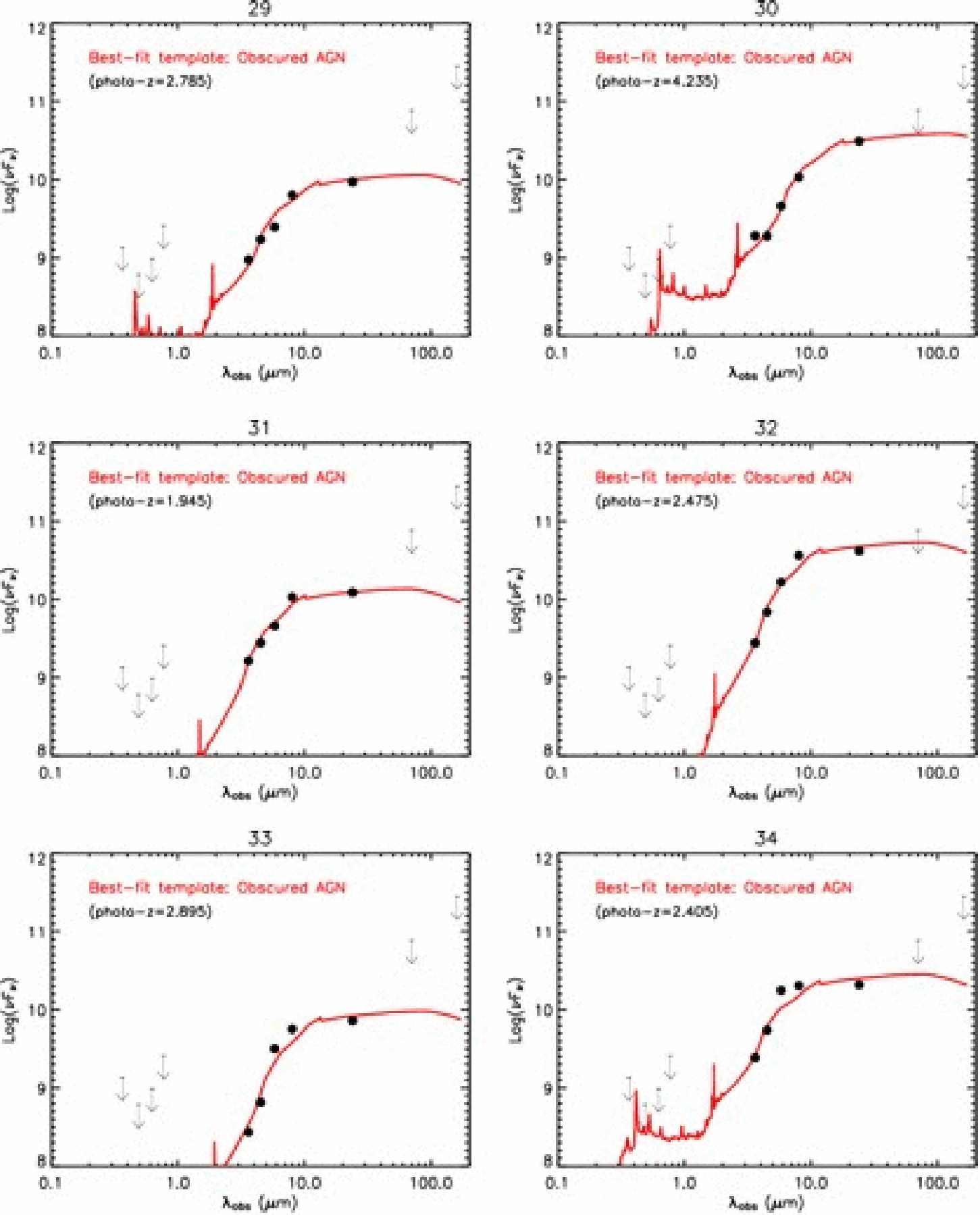}
\end{figure}

\clearpage

\section{Discussion and Conclusions}

We have presented a parametric model for describing SWIRE
galaxy populations and outliers in the fields of ELAIS-N1 and Lockman. 

For 16,698 sources in ELAIS-N1 with detections in four IRAC bands, we
use 3 IRAC colours and an optically derived photometric redshift and
find our data set is best described by four Gaussian modes.  ($C_{a}$,
$C_{b}$, $C_{c}$ and $C_{d}$).  We have determined the parameters of
these modes, providing an empirical description of this sub-set of the
SWIRE galaxies and shown (with a $\chi_{\nu}^2$ test) that our empirical
model of four Gaussian modes provides a good description of our SWIRE
data set.

We also find that by using only 3 IRAC colours (i.e. excluding photometric
redshift) our data set is still best modelled by the same four
Gaussian distributions. 

We then find that synthetic data from two theoretical models provide a
very poor description our empirical model. The model of Xu is found
to significantly underpredict the population of spiral galaxies with
infrared colour 0.2 $<$ log($f_{8.0}$/$f_{5.8}$) $<$ 0.6. The GalICS
model fails to account for AGN and ULIRGS redder than
log($f_{4.5}$/$f_{3.6}$) $>$ 0. This demonstrates that predictions from
these theoretical models are clearly inadequate for describing the
wealth of data in the SWIRE survey. When such models are available,
we have illustrated how our simple parametric description of the SWIRE
colour distribution can be used as a powerful model discriminator,
entirely complementary to comparisons of number counts.

We then use optical/infrared template colours and star formation
rate/stellar mass colour indicators to determine the galaxy types that
exist in each of our four distributions.
				   
Galaxies in population $C_{a}$ (magneta) are dusty systems with high
levels of star-formation activity such as ULIRGS, over a broad
redshift range.
Population $C_{b}$ (orange) contains low redshift galaxies dominated
by PAH emission at 8$\mu$m, characteristic of late-type spiral systems.  
Population $C_c$ (green) contain sources at intermediate
redshifts. The star-formation activity of this population is less than
that of the ULIRG population of $C_{a}$, but higher than that of the 
late-type spiral systems of population $C_{b}$, indicating that
$C_{c}$ is dominated by spiral and dusty starburst systems.  
Population $C_d$ (blue) is dominated by elliptical and early-type
spiral galaxies at low redshift ($\langle z_{ph d} \rangle$=0.32), 
which correspond to the bi-modality found using optical colours. Since
all our distributions were identified using IRAC colours which at
low redshift detect old stars where ellipticals and early-type spirals
are similar, the two galaxy classes are found within this single mode.  

We then devise a new technique for identifying unusual sources in the
fields of ELAIS-N1 and Lockman. We identify a total of 216 candidate
outliers. 

Analysing a selection of these outliers we find that sources with 
blue log($f_{4.5}$/$f_{3.6}$) colours are star-forming/Seyfert 2 systems
where sharp PAH features are responsible for their peculiar IRAC colours at 
particular redshifts. Outliers with red log($f_{4.5}$/$f_{3.6}$)
colour are found to be dusty star-forming systems such as
ULIRGs/Mrk231, where PAH features have thrown out the IRAC colours at
certain redshifts.

We also identify a sub-sample of 12 galaxies with very red infrared
SED's and no optical detections. Best modelled as obscured AGN at
redshifts of $z_{ph}$$\sim$ 2 -- 4, these sources would be very
infrared luminous, with $L_{ir}$$\sim$$10^{12.6-14.1}$$L_{\odot}$, and
would correspond to $\sim$0.1$\%$ of the number density of
sub-millimeter galaxies in the high redshift universe.  

We expect such extreme outliers to be dust enshrouded systems with a strongly
obscured AGN. The high mid-infrared emission may be as a result of dust being
heated by the AGN, similar to Compton-thick AGN (Polletta et al.\ 2006a). The hyperluminous gravitationally lensed galaxy IRAS F10214+4724 at z=2.29 (Rowan-Robinson et al.\ 1991 and Lacy et al.\ 1998) is known to contain dust-enshrouded AGN, and two ultraluminious high-redshift dusty galaxies H 1413+117 (Barvainis et al.\ 1995) and APM08279+5255 (Irwin et al.\ 1998) also contain powerful AGN that similarly dominate their infrared emission.
However, a starburst component in these galaxies cannot be completely ruled out. 
Therefore, an alternative interpretation could be that these systems
have high mid-infrared emission because they are going
through their maximal star formation whilst habouring AGN
activity (Lonsdale et al.\ 2006, in prep.). This present day level of 
star formation could be related to strong negative feedback effects 
limiting population III star formation during the earliest galaxy 
formation era (Sokasian et al.\ 2004). Therefore, we could now be seeing the first episodes of 
substantial star formation occurring in these individual galaxies. 
The presence of a starburst component in these galaxies would
correspond to infrared star-formation rates in the range 
1.5x10$^{3}$ - 5.2x10$^{4}$M$_{\odot}$/yr$^{-1}$.
Identifying such galaxies was a key goal of SWIRE. 
 
The classification technique we have used in this paper is an
efficient, automated way of identifying sub-samples of galaxies with
common photometric properties. An arbitrary number of input parameters
can be employed, giving the method great flexibility, particularly
when additional information such as photometric redshifts is
available. In addition, this technique is capable of classifying sources of
similar photometric properties without making any prior assumptions
about what these objects may be (q.v. SED template fitting).

Another important feature of the techniques we have used is the way we
 use multi-dimensional data.  Classical techniques use 2D projections
 e.g.  identification of AGN in infrared colour space (Lacy et al.\
 2004), simulation of the mid-infrared Spitzer colours (Sajina et al.\
 2005), and a colour-based classification of SWIRE sources using
 template libraries (Polletta et al.\ 2006b).  While these will be
 roughly consistent when applied to low dimensionality data, the
 limitations of projections will become more apparent as we move to
 higher dimensionality.  Further, we can classify objects using a
 range of colours in the optical, near, mid and far-infrared, include
 photometric redshifts, stellarity, morphology, and so create an
 ``overall classification'' based on more than the photometric
 properties of sources.
      
Fully automated techniques such as this and the
complementary SED template fitting method will be essential for
further analysis of the SWIRE fields.

\section*{Acknowledgements}

PD is supported by PPARC Studentship (PPA/S/S/2002/03500/), SJO is
supported by a Leverhulme Research Fellowship, SJO, IW and RSS are
supported by PPARC standard grant (PPA/G/S/2000/00508 \& PPA/G/S/2002/00481), TB is supported
by PPARC Studentship (PPA/S/S/2001/03217/). Optical data for this work
was provided using observations from the National Science Foundation's
(NSF) Mayall 4-meter Telescope at Kitt Peak National Observatory, near
Tucson, Arizona. The Expectation Maximisation code was supplied by 
``The Auton Lab', part of Carnegie Mellon University's School of 
Computer Science. Finally, we thank the anonymous referee for his/her comments. 

Support for this work, part of the Spitzer Space Telescope Legacy
Science Program, was provided by NASA through an award issued by the
Jet Propulsion Laboratory, California Institute of Technology under
NASA contract 1407.

\newpage 

\appendix

\section*{APPENDIX}

\vspace{1cm}

\setcounter{table}{5}

The $N$-dimensional Probability Density Function ($PDF$) is defined as
(see $\S$3.1):

\vspace{1.0cm}

$${PDF(\underline{x};\underline{\mu}_{i},\Sigma_{i})}= A.{\frac{1}{\sqrt{(2{\pi})^{N}{|}{\Sigma_{i}}{|}}}}{\exp}{\bigg[}{-}{\frac{1}{2}}({\underline{x}-\underline{\mu}_{i}})^T{\underline{\Sigma}_{i}}^{-1}({\underline{x}-\underline{\mu}_{i}}){\bigg]}$$

where:

\vspace{1cm}

\noindent{{$A$} represents the amplitude of each Gaussian

\noindent{\underline{$x$} represents the co-ordinates of each galaxy

\noindent{\underline{$\mu_{i}$} represents the mean co-ordinate of each Gaussian

\noindent{$N$ is the dimensions of the Gaussians

\noindent{and the covariance matrix $\Sigma_{i}$ of each distribution (\textit{i}) in
$N$-dimensional parameter space is defined as:}    

\vspace{1cm}

\[{\Sigma_{i}} =  \left( \begin{array}{cccc} {\sigma{_{1}}^{2}} &
  {\sigma{_{12}}} & .\hspace{0.1cm} . & {\sigma{_{1N}}} \\
  {\sigma{_{21}}} & {\sigma{_{2}}^{2}} & .\hspace{0.1cm} . &
  {\sigma{_{2N}}} \\ . & . & .\hspace{0.1cm} . & . \\ . & . &
  .\hspace{0.1cm} . & . \\ {\sigma{_{N1}}} & {\sigma{_{N2}}} &
  .\hspace{0.1cm} . & {\sigma{_{N}}^{2}}\end{array} \right) \]
 
where: 

\vspace{1cm}

\noindent{the variance $\sigma{_{N}}^{2}$ of each distribution is defined as $\sigma_{N}^{2} =  \langle({{x}_{N}-{\mu}_{N}})^2\rangle$}

\noindent{the covariance $\sigma{_{NM}}$ (N$\neq$M) is defined as $\sigma_{NM} =  \langle({{x}_{N}-{\mu}_{N}})({{x}_{M}-{\mu}_{M}})\rangle$}


\begin{deluxetable}{ccccc}
\tablecaption{\small{The amplitude ($A$), mean ($\mu$), variance $\sigma{_{N}}^{2}$
and covariance $\sigma{_{NM}}$ (N$\neq$M) of each distribution in
3-colour-plus-redshift space. Errors assigned to these values are
based on the variation between the ELAIS-N1 and Lockman data sets. The covariance matrix describing each distribution corresponds to galaxies above the SWIRE 5$\sigma$ flux limits - see $\S$2.1. }}
\tablewidth{0pt}
\tablehead{
\colhead{} & 
\colhead{$C_a$} & 
\colhead{$C_b$} &
\colhead{$C_c$} &
\colhead{$C_d$} \\
}
\startdata
$A$ & 0.11 $\pm$ 0.02 & 0.18 $\pm$ 0.06 & 0.25 $\pm$ 0.03 & 0.46 $\pm$ 0.03\\
$\mu_{1}$ & 0.13 $\pm$ 0.01 & 0.66 $\pm$ 0.06 & -0.05 $\pm$ 0.01 & 0.17 $\pm$ 0.05\\
$\mu_{2}$ & 0.12 $\pm$ 0.02 & -0.10 $\pm$ 0.02 & -0.02 $\pm$ 0.02 & -0.14 $\pm$ 0.01\\
$\mu_{3}$ & 0.07 $\pm$ 0.02 & -0.10 $\pm$ 0.03 & -0.16 $\pm$ 0.01 & -0.09 $\pm$ 0.01\\
$\mu_{4}$ & 1.28 $\pm$ 0.09 & 0.17 $\pm$ 0.03 & 0.73 $\pm$ 0.11 & 0.32 $\pm$ 0.04\\
$\sigma{_{1}}^{2}$ & 0.018 $\pm$ 0.003 & 0.033 $\pm$ 0.009 & 0.022 $\pm$ 0.006 & 0.064 $\pm$ 0.011\\ 
$\sigma{_{2}}^{2}$ & 0.017 $\pm$ 0.002 & 0.013 $\pm$ 0.001 & 0.011 $\pm$ 0.004 & 0.012 $\pm$ 0.003\\ 
$\sigma{_{3}}^{2}$ & 0.009 $\pm$ 0.002 & 0.005 $\pm$ 0.002 & 0.002 $\pm$ 0.002 & 0.003 $\pm$ 0.001\\
$\sigma{_{4}}^{2}$ & 0.336 $\pm$ 0.057 & 0.015 $\pm$ 0.005 & 0.052 $\pm$ 0.017 & 0.018 $\pm$ 0.011\\
$\sigma{_{12}}$ or $\sigma{_{21}}$ & 0.004 $\pm$ 0.002 & -0.003 $\pm$ 0.001 & -0.001 $\pm$ 0.003 & 0.004 $\pm$ 0.006\\
$\sigma{_{13}}$ or $\sigma{_{31}}$ & 0.005 $\pm$ 0.002 & -0.002 $\pm$ 0.002 & 0.003 $\pm$
0.001 & 0.009 $\pm$ 0.001\\
$\sigma{_{14}}$ or $\sigma{_{41}}$ & 0.003 $\pm$ 0.01 & -0.005 $\pm$ 0.007 & 0.002 $\pm$
0.003 & 0.007 $\pm$ 0.007\\
$\sigma{_{23}}$ or $\sigma{_{32}}$ & 0.006 $\pm$ 0.002 & -0.003 $\pm$ 0.0005 & 0.0004
$\pm$ 0.0007 & 0.002 $\pm$ 0.001\\
$\sigma{_{24}}$ or $\sigma{_{42}}$ & 0.005 $\pm$ 0.012 & -0.004 $\pm$ 0.0003 & -0.003
$\pm$ 0.008 & 0.006 $\pm$ 0.004\\
$\sigma{_{34}}$ or $\sigma{_{43}}$ & 0.013 $\pm$ 0.006 & 0.006 $\pm$ 0.002 & 0.0007 $\pm$
0.002 & 0.003 $\pm$ 0.002\\
\enddata
\medskip

\footnotesize{1=log($f_{8}$/$f_{5.8}$), 2=log($f_{5.8}$/$f_{4.5}$),
3=log($f_{4.5}$/$f_{3.6}$) and 4=photometric redshift}   
\end{deluxetable}

\label{lastpage}

\end{document}